\documentclass[aip,reprint,superscriptaddress]{revtex4-2}

\usepackage{amsmath,amsfonts,amssymb,mathrsfs,bm}
\usepackage{marginnote}
\usepackage{gensymb} 
\usepackage{graphicx}
\usepackage{placeins}
\usepackage[colorlinks]{hyperref}
\usepackage[dvipsnames]{xcolor}
\usepackage{oplotsymbl}
\usepackage[normalem]{ulem}

\definecolor{myblue}{RGB}{0,76,153}
\definecolor{mycite}{RGB}{0,170,0}
\hypersetup{
    colorlinks=true,
    linkcolor=myblue,
    filecolor=myblue,      
    urlcolor=myblue,
    citecolor=mycite
}

\DeclareMathOperator{\tr}{\textrm{tr}}
\renewcommand{\vec}[1]{\boldsymbol{#1}}
\newcommand{\tens}[1]{\mathbf{#1}}

\newcommand{\D}{\textrm{D}}

\newcommand{\Wi}{\textrm{Wi}}
\newcommand{\bnabla}{\vec{\nabla}}

\newcommand{\gammadot}{\dot{\gamma}}

\newcommand{\Zrhe}{Z} 
\newcommand{\Zrhec}{Z_{c}} 
\newcommand{\lmax}{\lambda_{\textrm{max}}}
\newcommand{\tauR}{\tau_{\textrm{R}}}

\newcommand{\WiR}{\Wi_\textrm{R}}
\newcommand{\lj}{\scriptscriptstyle{\textrm{LJ}}}
\newcommand{\pp}{\textrm{pp}}
\newcommand{\eq}{\textrm{eq}}

\graphicspath{{./Figures/}}

\begin{document}
	
	\title{Shear Banding in Simulations of Polymer Melts}
	
	\author{Lucas L. Nelson}
	\email{lln20@georgetown.edu}
    \affiliation{Georgetown University, Department of Physics and Institute for Soft Matter Synthesis and Metrology, Washington DC, 20057 USA}
        \author{Gary S. Grest}
    \email{gsgrest@sandia.gov}
	\affiliation{Sandia National Laboratories, Albuquerque, New Mexico 87185, USA}
    \author{Peter D. Olmsted}
	\email{pdo7@georgetown.edu}
	\affiliation{Georgetown University, Department of Physics and Institute for Soft Matter Synthesis and Metrology, Washington DC, 20057 USA}
	
	\date{\today}
	
	\begin{abstract}
		Results from numerical simulations of polymers under shear flow are compared with predictions for shear banding based on a model coupling Rolie-Poly-like tube dynamics to the entanglement dynamics mediated by Convected Constraint Release (CCR).
		CCR is controlled by a parameter $\beta$, whose dependence on the bending stiffness $k_{\theta}$ is calculated from the simulations.
		The model predicts shear banding for polymers whose equilibrium entanglement number $\Zrhe$ exceeds a critical value $\Zrhec$ that depends on $\beta$.
		The simulations are in semi-quantitative agreement with the model, with deviations that are attributed to approximations inherent in the model, and the inadequacy of tube models to describe partially-disentangled liquids in strong flow.
        These results may help determine which physical polymers could undergo shear banding.
	\end{abstract}
	
	
	\maketitle

\label{sec:intro}
\section{Introduction}
Polymeric liquids are strongly entangled  when the polymer molecular weight $M$ greatly exceeds a critical value $M_c$ \cite{DoiEdwards_Book,de_gennes_reptation_1971}.
The dynamics of entangled polymers are well-described by tube models such as the Doi-Edwards (DE) model \cite{DoiEdwards_Book,glmm-03,milner-20}, in which polymers diffuse within a `tube' of $\Zrhe$ distinct entanglements, and reset their conformation on the scale of the terminal `reptation' time $\tau_d \sim 3 \Zrhe^3$. 
The DE model predicts a non-monotonic steady state relation between shear stress $\sigma_{xy}$ and shear rate $\gammadot$ with a stress maximum when $\gammadot \simeq 1/\tau_d$. This can lead to instabilities such as spurt from a pipe \cite{vinogradov_viscoelastic_1972,mcleish86} and shear banding \cite{hunter1983vff,pearson94,spenley93,spenley96,olmsted99a,olmstedbanding08}, whose simplest signature is a steady state constant `stress plateau' in the shear stress as a function of shear rate beginning around $\gammadot \simeq 1/\tau_d$.

Most experiments indicate a slight increase in stress rather than a plateau. \cite{burroughs_flow-induced_2021} To eliminate this inconsistency in DE theory, Marrucci and Ianniruberto \cite{marrucci-96,im-14,im-14b,ianniruberto-15}
incorporated polymer stretch and its relaxation due to non-equilibrium \textit{convected constraint release} (CCR), which speeds up the terminal time.
In CCR, polymers stretched by flow retract more quickly, which accelerates equilibrium (thermal) constraint release, leads to a more disordered tube contour, and increases the shear stress.
CCR is controlled by a parameter $\beta$: for large enough $\beta$ the increased stress eliminates non-monotonicity and thus eliminates shear banding \cite{glmm-03,adams11RP}.

Some highly entangled polymer solutions demonstrate phenomena consistent with non-monotonic constitutive curves, including discontinuous shear rate jumps in stress-driven flow \cite{tapadiawang03} and banded velocity profiles in startup \cite{ravindranath2008banding}, large-amplitude oscillatory \cite{li_elastic_2010}, and steady-state \cite{tapadiawang06,boukany_shear_2015} shear flows.
These results are debated due the confounding factors of wall slip \cite{BlaGra1999PF,BlaGra1996PRL} and edge fracture \cite{hemingway2018edge,hemingway2020interplay}, with some studies with similar materials and conditions not observing banding \cite{inn2005eef,li2013flow,li_startup_2015}.
This may be because the formation of shear bands 
is history-dependent \cite{nohel90,malkus90,grand_slow_1997,olmstedbanding08, hu_steady-state_2010,cheng2012shear}.
\citet{wang_wall_2019} attributed flow inhomogeneities to disentanglement in the higher-shear-rate band,
while other studies \cite{fielding03b,fielding03c, cromer_shear_2013,burroughs_flow-induced_2021,burroughs_flow-concentration_2022} have coupled shear banding to concentration fluctuations.

We are not aware of incontrovertible experimental evidence for shear banding in polymer melts. Observations such as the spurt effect \cite{vinogradov_viscoelastic_1972} and long stress plateaus \cite{auhl2008linear} suggest banding-like behavior, and phenomena such as transient banding \cite{sun_shear_2012} and other flow inhomogeneities \cite{fang_shear_2011, li_strain_2018}  are again complicated by wall slip \cite{BlaGra1999PF,BlaGra1996PRL} and edge fracture \cite{hemingway2018edge,hemingway2020interplay},  which makes steady-state observations difficult at high $\Zrhe$ \cite{li_practical_2024}.
By contrast, simulations show  clear evidence for steady-state banding in polymer melts.
\citet{CaoLikhtmanPRL2012}, \citet{mk16model,mk16banding}, and \citet{ruan_shear_2021} observed shear banding in bead-spring polymer simulations of steady-state shear.
Most simulations of polymers under shear use the SLLOD algorithm \cite{evans1984nonlinear}, which imposes a uniform shear flow and does not allow shear banding \cite{CaoLikhtmanPRL2012}.

To understand these phenomena, we consider flow-induced disentanglement, which is encapsulated in CCR.
The number of entanglement ``kinks'' $Z_k$ measured by primitive path analysis \cite{everaers_rheology_2004, kroger23Z1} has been shown in simulations \cite{bmk-10, xu_influence_2020, ruan_shear_2021, gor-22, dolata24chemical} to decrease in strong flows with $\gammadot\gtrsim 1/\tauR$, where $\tauR\sim \Zrhe^2$ is the Rouse time, agreeing with CCR predictions.
To model this behavior, \citet{im-14} applied the CCR mechanism to the Doi-Edwards model and derived equations of motion coupling the reduced number of entanglements $\nu(t)\equiv Z_k(t)/Z_{k,\eq}$ to the polymer orientation tensor, reproducing the steady state reduction of entanglements $\nu(\gammadot)$ found in simulations by \citet{bmk-10}. More recently, \citet{dolata22thermo} (DO) derived a thermodynamically-consistent version of this model coupling CCR to the Rolie-Poly model, and showed, by comparing with simulations, that the CCR parameter $\beta$ is larger for stiffer polymers \cite{dolata24chemical}.

In this paper we compare predictions of the DO model for shear banding with observations of shear banding in Kremer-Grest bead-spring simulations \cite{kg-90}.
The DO model predicts banding for sufficiently-entangled polymers $\Zrhe>\Zrhec(\beta,\alpha)$, where the critical entanglement number $\Zrhec$ is larger for larger CCR parameter $\beta$ and smaller 
relaxation anisotropy $\alpha$ (defined below).
The region of banding found in the simulations agrees surprisingly well with the model, given that the DO (and Rolie-Poly) model is a simplified single-mode version of the more precise microscopic GLaMM model \cite{glmm-03}.
We speculate that physical polymers whose dynamics are controlled by dihedral rotations have larger $\beta$, which may explain why shear banding is not seen in polymer melts.

\section{Theory}
\subsection{DO Model}
The DO model couples the evolution of the local tube conformation tensor $\tens{A}(t)$, roughly following the Rolie-Poly (RP) model \cite{likhtmangraham03}, to the reduced number of primitive-path kinks $\nu(t) = Z_k(t)/Z_{k,\eq}$.
These quantities, and the local stress tensor $\vec{\sigma}$, evolve according to the following equations:

\begin{subequations}\label{eq:governing-equations}
    \begin{align}
        \begin{split}
			\stackrel{\triangledown}{\tens{A}}  & = 
				- \frac{\tens{I} + \alpha(\tens{A}-\tens{I})}{\tilde{\tau}_d(\lambda, \beta)}\cdot(f(\lambda)\tens{A} - \tens{I}) 
				\\&- \frac{2}{\tauR} \frac{f(\lambda) \lambda^2 - 1}{\lambda^2 + \lambda}\tens{A} \\&- \frac{\zeta_Z\beta\nu}{3\lambda^2}
			\left(\frac{\tens{I} + \alpha(\tens{A} - \tens{I})}{\tilde{\tau}_d(\lambda,\beta)}
				+ \frac{2\lambda \tens{I}}{(\lambda + 1)\tauR}\right)\cdot \tens{A}\ln\nu,
	\end{split}\label{eq:governing-equation-A}\\
	\begin{split}
		\frac{\D \nu}{\D t} & = -\frac{\beta\nu}{3\lambda^2}\left( \tens{A}:\bnabla \vec{v} - \frac{1}{2}\frac{\D\tr\tens{A}}{\D t}\right) 	- \frac{\ln\nu}{\tauR},		
	\end{split}\label{eq:governing-equation-nu} \\
	\vec{\sigma} & = G_0\left(f(\lambda)\tens{A} - \tens{I}\right) + 2\eta \tens{D}. \label{eq:stress}
	\end{align}
\end{subequations}

Here, $\stackrel{\triangledown}{\tens{A}}$ is the upper convected derivative
\begin{equation}
	\stackrel{\triangledown}{\tens{A}} \equiv \frac{\D\tens{A}}{\D t} - \tens{A}\cdot\bnabla\vec{v} - \left( \bnabla\vec{v} \right)^{\text{T}} \cdot\tens{A};
\end{equation}
$\tilde{\tau}_{d}(\lambda, \beta)$ is a non-equilibrium reptation time given by
\begin{equation} \label{eq:taud}
    \frac{1}{\tilde{\tau}_d(\lambda, \beta)} \equiv \frac{1}{\tau_{d}} +
    \frac{Z_{k,\eq}}{\nu Z_{k,\eq}+1}\frac{\beta\nu}{\tauR\lambda}\frac{f(\lambda)\lambda^2 - 1}{\lambda+1};
\end{equation}
$\zeta_Z \equiv Z_{k,\eq}/\Zrhe$ is the ratio of the number of kinks to the standard entanglement number $\Zrhe$;
$\lambda \equiv \sqrt{\tr \tens{A}/3}$ is the relative stretch of the polymer;
and $f(\lambda)$ is the Cohen approximation~\cite{cohen-91} to the Finitely-Extensible Nonlinear Elastic (FENE) spring force
\begin{equation} \label{eq:fene}
f(\lambda) = 1 + \frac{2}{3} \frac{\lambda^2 - 1}{\lmax^2 - \lambda^2}.
\end{equation}
The  maximum relative stretch $\lmax \equiv \sqrt{N_{e\textrm{K}}}$, where $N_{e\textrm{K}}$ is the number of Kuhn steps per entanglement.
This term diverges as $\lambda\to\lmax$, but in the limit $\lmax\to\infty$, $f(\lambda)=1$, and the behavior of the chain becomes Gaussian.

Previous studies indicate that $Z_{k,\eq} \approx 2\Zrhe$ \cite{bmk-10}, where $\Zrhe$ is derived from the linear modulus $G_0$ \cite{DoiEdwards_Book}; the DO model assumes an exact relation $Z_{k,\eq} = 2 \Zrhe-1$ \cite{dolata22thermo}.
The relation between the Rouse time $\tauR$ and the (equilibrium) reptation time $\tau_{d}$ is given by
\begin{gather} \label{eq:timescales}
   \frac{\tau_{d}}{\tauR} = 3\Zrhe f_L(\Zrhe), \\
   \noalign{\noindent where the Likhtman relation}
   f_L(\Zrhe) =1 - {3.38}{\Zrhe^{-1/2}}+{4.17}{\Zrhe^{-1}} - {1.55}{\Zrhe^{-3/2}} \label{eq:likhtman}
\end{gather}
incorporates equilibrium constraint release and contour length fluctuations \cite{lm02}. 

The CCR parameter $\beta$ is a positive parameter of order 1, which determines the strength of convective constraint release.
The relaxation anisotropy parameter $\alpha$ controls a Giesekus-like term in the model \cite{giesekus-82}, which allows for nonzero second normal stress differences not permitted by the RP model.
Its value should lie between 0 (for isotropic relaxation) and 1 (for maximum anisotropy), and it determines the ratio of the normal stress differences in the low-shear-rate limit: $\alpha = -2 \lim_{\dot{\gamma}\to 0}(N_2/N_1)$\cite{dolata22thermo}, where $N_1$ and $N_2$ are the first and second normal stress differences.
Existing data on polymer solutions indicate $\alpha \approx 0.6$ \cite{keentok_coneplate_1982,ramachandran_dependence_1985,magda_concentrated_1994}, while data on melts suggest $\alpha>0.5$ \cite{schweizer2002measurement,schweizer2004nonlinear,schweizer_shear_2008,schweizer_cone-partitioned_2013,costanzo_measuring_2018,li_normal_2025} but is not conclusive.
We discuss this issue in more detail in Section~\ref{sec:normal_stress}.

To account for fast fluctuations at scales smaller than an entanglement,
we add a Newtonian stress $2\eta\tens{D}$ to the stress term from the original DO model,
where $\eta$ is a background viscosity and $\tens{D} = \frac{1}{2}\left[ \bnabla \vec{v} + (\bnabla \vec{v})^T \right]$ is the fluid's symmetric strain rate tensor. Following \citet{adams11RP}, we parameterize this background viscosity by $\varepsilon = {\eta}/\eta_p$, where $\eta_p=G_0\tau_{d}$ is the low-shear-rate viscosity of the entangled liquid.

For melts, the background viscosity is $\eta_e \equiv c \zeta N_{e\textrm{K}} b_{\textrm{K}}^2/36$, which is the (Rouse) viscosity for shorter chains of $N_{e\textrm{K}}$ Kuhn steps of length $b_{\textrm{K}}$ with Kuhn segment number density $c$ and monomer friction constant $\zeta$ \cite{DoiEdwards_Book}.
The plateau modulus of the full chains is $G_0=c k_B T/N_{e\textrm{K}}$ \cite{DoiEdwards_Book}, so the viscosity ratio becomes
\begin{equation} \label{eq:melt_eps}
\varepsilon_{\textrm{melt}} =
\frac{\eta_e}{\eta_p} =
\frac{\pi^2}{36\Zrhe^3 f_L(\Zrhe)}.
\end{equation}
For dilute solutions, $\eta$ is the solvent viscosity, so $\varepsilon$ no longer has this exact form, but  still scales as $\varepsilon\sim\eta_s\Zrhe^{-3}$. 

The DO model has five dimensionless parameters: $\Zrhe$, $\beta$, $\alpha$, $\lmax$, and $\varepsilon$. For melts, $\varepsilon$ is determined by $\Zrhe$, while $\Zrhe,\lmax$, and $\alpha$ can be determined from linear or weakly-nonlinear rheometry along with the modulus and time scales. This leaves $\beta$ as, strictly, the only free parameter. 

\subsection{Predictions of Banding}\label{sec:analytical}

\begin{figure}[htp]
    \centering
    \includegraphics[width=1\linewidth]{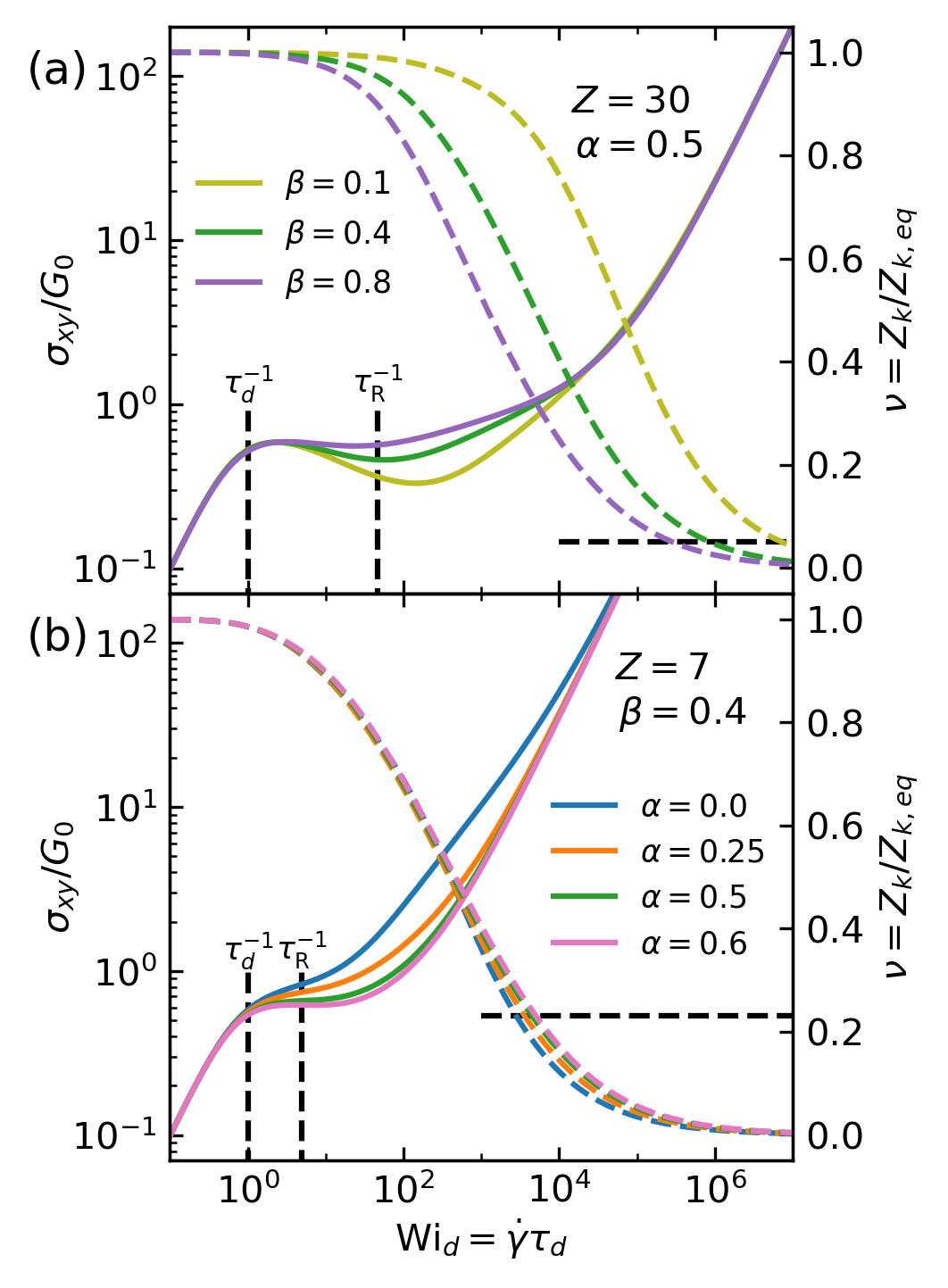}
    \caption{Stress-strain rate curves (solid lines) and relative entanglement (dashed lines) vs reptation Weissenberg number for the DO model with $\Zrhe=30$ (a) or $7$ (b), varying $\beta$ (a) or $\alpha$ (b) with $\lmax=4$, $\varepsilon=\varepsilon_{\textrm{melt}}$. Vertical dashed lines indicate the reciprocal Rouse and reptation timess, and horizontal dashed lines indicate the value of $\nu$ for which $Z_k=3$; when the number of entanglements drops below this value, the tube model is likely to break down}
    \label{fig:stress_strainrate}
\end{figure}

To determine when the model predicts banding, we numerically solve for $\nu$ and $\tens{A}$ in steady state homogeneous simple shear flow to calculate the shear stress $\sigma_{xy}$ as a function of the shear rate $\gammadot$ (Fig.~\ref{fig:stress_strainrate}).
We express this shear rate in terms of two different Weissenberg numbers in this paper: the reptation Weissenberg number $\Wi_d = \gammadot \tau_d$ based on the terminal relaxation time, and the Rouse Weissenberg number $\WiR=\gammadot\tauR$ which specifically governs disentanglement.
These constitutive curves are monotonic for some parameter combinations, but become non-monotonic for $1/\tau_{\textrm{d,\eq}}\lesssim\dot{\gamma}\lesssim 1/\tauR$ for others, which coincides with a reduction in entanglements $\nu(\Wi_d)<1$. 
When $Z_k=\nu Z_{k,\eq}$ drops below a value of about 3 (horizontal dashed lines in Fig.~\ref{fig:stress_strainrate}), the concept of an entanglement tube becomes poorly-defined and we expect the model to fail.

\begin{figure}[htp]
    \centering
    \includegraphics[width=0.9\linewidth]{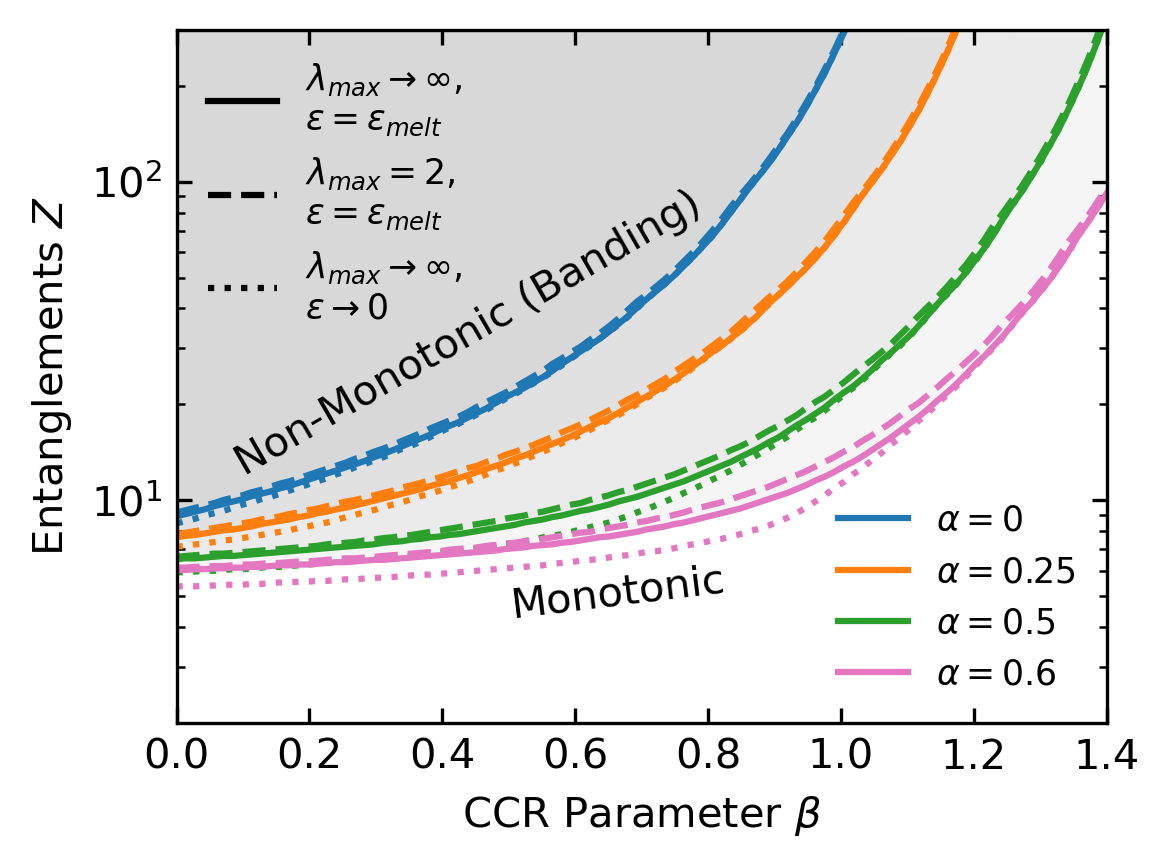}
    \caption{Boundaries $\Zrhec(\beta,\alpha)$ between DO model parameters with non-monotonic (shaded) and monotonic (unshaded) constitutive curves, for various values of $\alpha$, $\varepsilon$, and $\lmax$.}
    \label{fig:phase_diagram}
\end{figure}

We use these constitutive curves to construct a ``phase diagram'' for monotonicity, and thus for banding (Fig.~\ref{fig:phase_diagram}).
The monotonicity boundary depends weakly on $\lmax$ and $\varepsilon$, even for extreme variations $\lmax\in [2,\infty)$ (solid vs. dashed lines) and $\varepsilon \in (0,\varepsilon_{\textrm{melt}}]$ (solid vs. dotted) (see Appendix~\ref{app:eps_lmax} for a discussion).

Increasing $\Zrhe$ promotes banding \cite{DoiEdwards_Book,ravindranath2008banding,adams11RP} by increasing the separation between the Rouse and reptation times, broadening the range of shear rates in which the stress can decrease (compare the green curves in Figs.~\ref{fig:stress_strainrate}ab, for which all other parameters are identical).
Strengthening CCR (increasing $\beta$) inhibits banding, as proposed by \citet{im-96}. Fig.~\ref{fig:stress_strainrate}a demonstrates the mechanism: stronger CCR
increases disentanglement (smaller $\nu$) when $\gammadot \sim 1/\tauR$, which decreases polymer alignment and increases $\sigma_{xy}$.
This value of $\gammadot$ is typically near the local stress minimum, so a sufficient increase of $\sigma_{xy}$ there can eliminate the nonmonotonicity.

Significantly, a larger relaxation anisotropy $\alpha$ promotes banding.
As evident in Fig.~\ref{fig:stress_strainrate}b, increasing $\alpha$ decreases the shear stress for a range of shear rates above $1/\tau_d$, which promotes nononotonicity.
Mechanically, increasing $\alpha$ accelerates the relaxation of $\tens{A}$ along directions in which it is elongated, so the polymer component of $\sigma_{xy}$ decays more quickly as $\gammadot$ increases.

\section{Model and Methodology}\label{sec:simulations}
To test these predictions, we performed non-equilibrium molecular dynamics simulations of semi-flexible Kremer-Grest (KG) bead-spring polymers~\cite{kg-90} using the large scale atomic molecular massively parallel simulator (LAMMPS) software \cite{thompson2022lammps}.
The model consists of beads of mass $m$ and diameter $\sigma$ connected with FENE bonds, which interact with a Lennard-Jones potential with interaction strength $\epsilon_{\lj}$ cut off at $r_c= 2^{1/6}\sigma$. To vary the entanglement length, a 3-body cosine angle interaction $k_\theta \epsilon_{\lj} \left[1+\cos(\theta) \right]$, with $k_\theta=0,\, 1.5, $ and $2.0$ was included.
Chains lengths $N_{\textrm{mon}}=200$ to $2000$ were studied, depending on the bending stiffness. The number density is $0.85 \sigma^{-3}$ for all systems.

Most of the systems are in a cubic box with periodic boundary conditions, with equilibrium configurations prepared following the procedure described \citet{auhl_equilibration_2003}.
To determine $\beta$, some systems were then replicated in the $x$ direction four times and equilibrated using the double bridging algorithm \cite{auhl_equilibration_2003} before applying shear.
See Appendix~\ref{app:simsmore} for a list of all systems studied.

The nonequilibrium MD simulations were performed using two different procedures.
The first uses the SLLOD equations of motion \cite{evans1984nonlinear,daivis2006simple,baig2005proper,edwards2006validation} with a Nose-Hoover thermostat \cite{nose1984unified,hoover1985canonical} and a temperature damping
constant of $\tau \equiv \sqrt{m\sigma^2/\epsilon_{\lj}}$.
This algorithm imposes a linear velocity profile in the shear plane and thus prevents any shear banding. 
In the second procedure, the simulation cell is deformed at a constant rate in the shear plane using the {\tt fix deform} command in LAMMPS, and the beads are coupled to a pairwise DPD thermostat \cite{hoogerbrugge1992simulating,espanol1995hydrodynamics,soddemann2003dissipative} to control the temperature.
This thermostat acts only between neighboring beads with a distance $r_c$ and damping coefficient $\gamma=m/\tau$. As it does not impose a uniform velocity profile, this procedure allows shear banding. Both procedures give the same shear stress under all conditions, even when the system forms shear bands using the second method while the first method supresses the shear bands.

Due to computational limitations arising
from the need to measure time scales ranging from bond
relaxation times to that of the entire polymer chain, the strain rates $\dot{\gamma}$ currently
accessible in this study are $10^{-6}$ to $10^{-2}/\tau$.
The time step for all shear simulations was $0.008-0.01\tau$, and all simulations were performed at constant volume.
We compute the entanglement Rouse time $\tau_e$, number of Kuhn steps per entanglement $N_{e\text{K}}$, and ratio of Kuhn step to packing length $b_{\text{K}}/p$ for these systems using formulas from \citet{everaers20KG}, then compute $\tau_d$ and $\tauR$ using Equation~\ref{eq:timescales} and the relation $\tauR = \Zrhe^2 \tau_e$; these quantities are given in Table~\ref{table:sims} and Table~\ref{table:simsmore}.
We determine $\Zrhe$ from equilibrium simulations by $\Zrhe=\langle L_{\pp} \rangle^2/\langle R^2 \rangle$, where $\langle R^2 \rangle$ is the mean squared end-to-end length and $\langle L_{\pp} \rangle$ is the mean primitive path length measured by the Z1+ code \cite{kroger23Z1}.

\begin{figure}[htb!]
    \centering
    \includegraphics[width=1\linewidth]{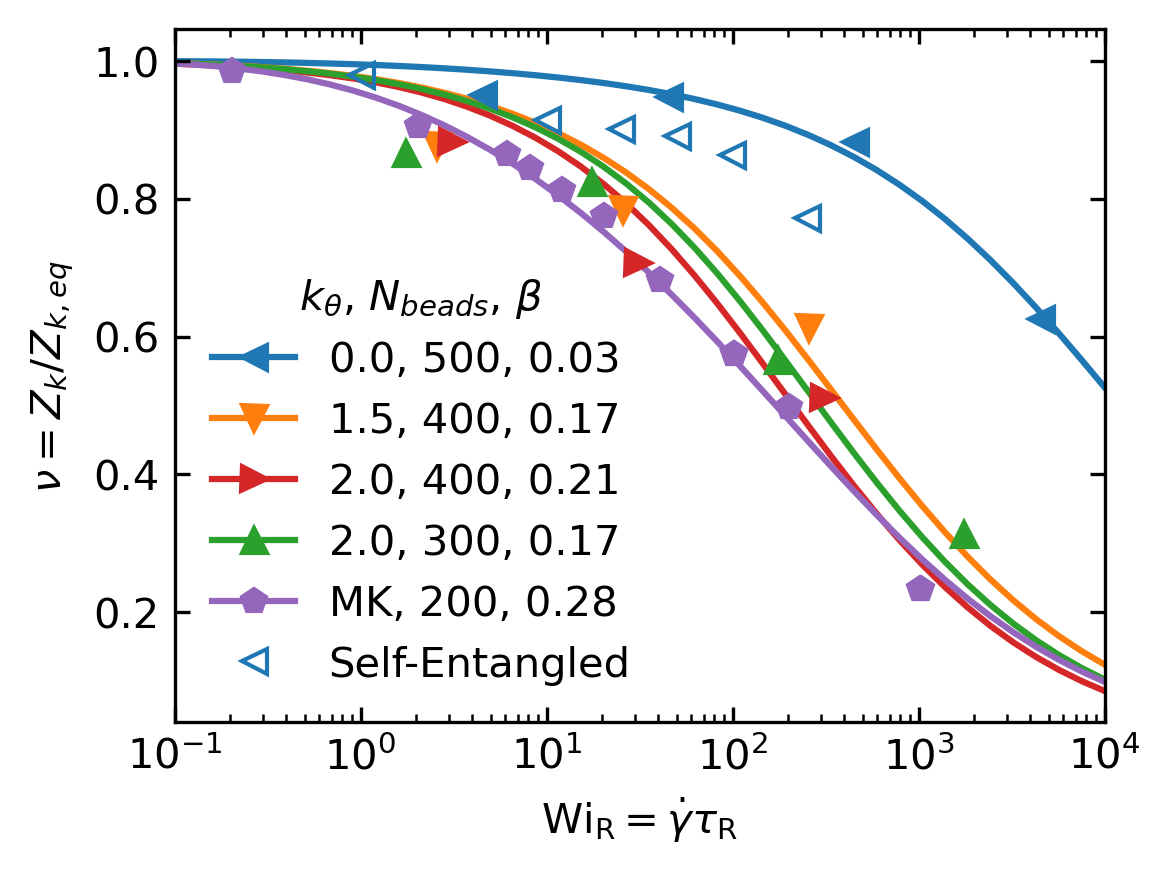}
    \caption{Disentanglement as a function of Rouse Weissenberg number for four KG systems (triangles) and simulations by \citet{mk16model} (pentagons).
	Lines are least-squares fits of the DO model to each of these systems, with the fit value of $\beta$ given.
	Hollow triangles are data from \citet{dolata24chemical} computed using the Z1 code on a smaller system susceptible to self-entanglement due to periodic boundary wrapping.}
    \label{fig:beta_fit}
\end{figure}

To measure disentanglement $\nu=Z_k/Z_{k,\eq}$ as a function of shear rate, we analyzed the simulations using the Z1+ code~\cite{kroger23Z1} to compute $Z_k$.
The Rouse Weissenberg number $\WiR = \gammadot\tauR$ at which $Z_k$ begins to drop below its equilibrium value in the DO model depends strongly on $\beta$, so by fitting its prediction of $\nu$ vs $\WiR$ to our simulations, as shown in Fig.~\ref{fig:beta_fit}, we are able to determine appropriate values of $\beta$.
The $\nu$ vs $\WiR$ curves we calculate assuming non-shear banded flow, so we perform this fitting on systems using the SLLOD shearing procedure, however measuring disentanglement on DPD simulations which do not shear band gives very similar results.
In this fitting process, we set $\lmax=\sqrt{N_{e\text{K}}}$ in the DO model and arbitrarily choose $\alpha=0.5$, as the latter has little effect on $\nu$.

This technique of determining $\beta$ was introduced by \citet{dolata24chemical}, using the earlier Z1 code \cite{karayiannis_combined_2009}.
However, the simulation sizes used by Dolata \textit{et al.} were small enough that at high shear rates, the flow-aligned polymer chains would wrap around the periodic boundary conditions and interact with periodic copies of themselves.
This artificially decreased the value of $Z_k$ measured by Z1 and Z1+, which by default do not count kinks which result from a chain interacting with itself; the hollow symbols in Fig.~\ref{fig:beta_fit} are data from Dolata \textit{et al.} and show the magnitude of this increased disentanglement when compared with the new, larger systems.
Z1+ is less susceptible to this issue than Z1, and replicating the system in the shear direction before running Z1+ mostly mitigates the issue, but to be certain we only perform our $\beta$ fitting on simulations which are longer in the shear direction than the maximally-extended chain length.

As a result of this discrepancy, the values of $\beta$ we find for our simulations are somewhat lower than reported in \citet{dolata24chemical} for analogous systems, but the main trends reported still hold: $\beta$ is greater for chains with greater stiffness $k_\theta$, but does not substantially depend on chain length for systems of the same stiffness (compare $k_\theta=2.0$ with chains of length 300 and 400  beads in Fig.~\ref{fig:beta_fit}, which agree within error bars).
Consequently, we assume $\beta$ to be constant for any given $k_\theta$, including for chain lengths whose disentanglement we have not measured.

To supplement our own simulations, we use the same method to fit $\beta$ in simulations by \citet{mk16model,mk16banding} (MK), who investigated banding in a slightly different bead-spring model with three different lengths of chain, and also measured disentanglement using the Z1 code.
Their simulations were all performed using a DPD procedure, and two of the three lengths exhibited banding, so we used their reported $Z_k$ vs $\WiR$ data for the non-banding system to fit $\beta$.

\begin{table}[htb]
\begin{tabular}{llllllcll}
\hline \hline
Model & Ref.                           & $k_\theta$ & $N_{\text{mon}}$ & $\Zrhe$ & $\beta$         & Bands & $b_{\textrm{K}}/p$ & $N_{e\textrm{K}}$ \\ \hline
F-KG  & \cite{dolata24chemical},$\dag$ & 0          & 500$^*$          & 6       & $0.03 \pm 0.01$ & N     & 2.8                & 39                \\
F-KG  & $\dag$                         & 0          & 1000             & 13      & $0.03 \pm 0.01$ & N     & 2.8                & 39                \\
F-KG  & $\dag$                         & 0          & 1500             & 19      & $0.03 \pm 0.01$ & Y     & 2.8                & 39                \\
F-KG  & $\dag$                         & 0          & 2000             & 26      & $0.03 \pm 0.01$ & Y     & 2.8                & 39                \\
SF-KG & $\dag$                         & 1.5        & 200              & 6       & $0.17 \pm 0.08$ & N     & 6.8                & 8.8               \\
SF-KG & $\dag$                         & 1.5        & 300              & 9       & $0.17 \pm 0.08$ & N     & 6.8                & 8.8               \\
SF-KG & $\dag$                         & 1.5        & 400$^*$          & 12      & $0.17 \pm 0.08$ & Y     & 6.8                & 8.8               \\
SF-KG & $\dag$                         & 2          & 200              & 8       & $0.19 \pm 0.06$ & N     & 9.9                & 5.5               \\
SF-KG & \cite{ruan_shear_2021},$\dag$  & 2          & 300$^*$          & 12      & $0.19 \pm 0.06$ & Y     & 9.9                & 5.5               \\
SF-KG & $\dag$                         & 2          & 400              & 16      & $0.19 \pm 0.06$ & Y     & 9.9                & 5.5               \\
MK    & \cite{mk16banding}             & 2          & 200$^*$          & 7       & $0.28 \pm 0.02$ & N     & 11.2               & 7.9               \\
MK    & \cite{mk16banding}             & 2          & 250              & 9       & $0.28 \pm 0.02$ & Y     & 11.2               & 7.7               \\
MK    & \cite{mk16banding}             & 2          & 400              & 14      & $0.28 \pm 0.02$ & Y     & 11                 & 8                 \\
UA-PE & \cite{sek-15}                  & --         & 400              & 5       & $1.25 \pm 0.11$ & --    & 12                 & 6                 \\
UA-PE & \cite{sek-16}                  & --         & 700              & 8.7     & $1.11 \pm 0.08$ & --    & 12                 & 6.2               \\
UA-PE & \cite{sek-19}                  & --         & 1000             & 12.9    & $0.98 \pm 0.16$ & --    & 10                 & 6.4               \\
\hline\hline
\end{tabular}
\caption{Simulation parameters (where $\dag$ indicates our results): bending stiffness $k_\theta$, number of beads $N_{\textrm{mon}}$, number of rheological entanglements $\Zrhe$ and fitted CCR parameter $\beta$; whether the system bands; the ratio of Kuhn and packing lengths $b_{\textrm{K}}/p$; and the numbers of Kuhn steps per entanglement $N_{e\textrm{K}}$. UA-PE data reproduced from \cite{dolata24chemical}. Asterisks indicate chain lengths used to fit $\beta$ and measure $\Zrhe$ with the Z1+ code.}
\label{table:sims}
\end{table}

Table~\ref{table:sims} collects the fit values of $\beta$ for the KG systems studied and the three systems studied by \citet{mk16model}, along with the estimated values of $\Zrhe$ and whether they exhibit banding.
The normal stress ratio $-\alpha/2$ cannot be determined with confidence, but is independent of $\Zrhe$ and $k_{\theta}$, as discussed in Section~\ref{sec:normal_stress}.

\begin{figure}[htb!]
    \centering
    \includegraphics[width=1\linewidth]{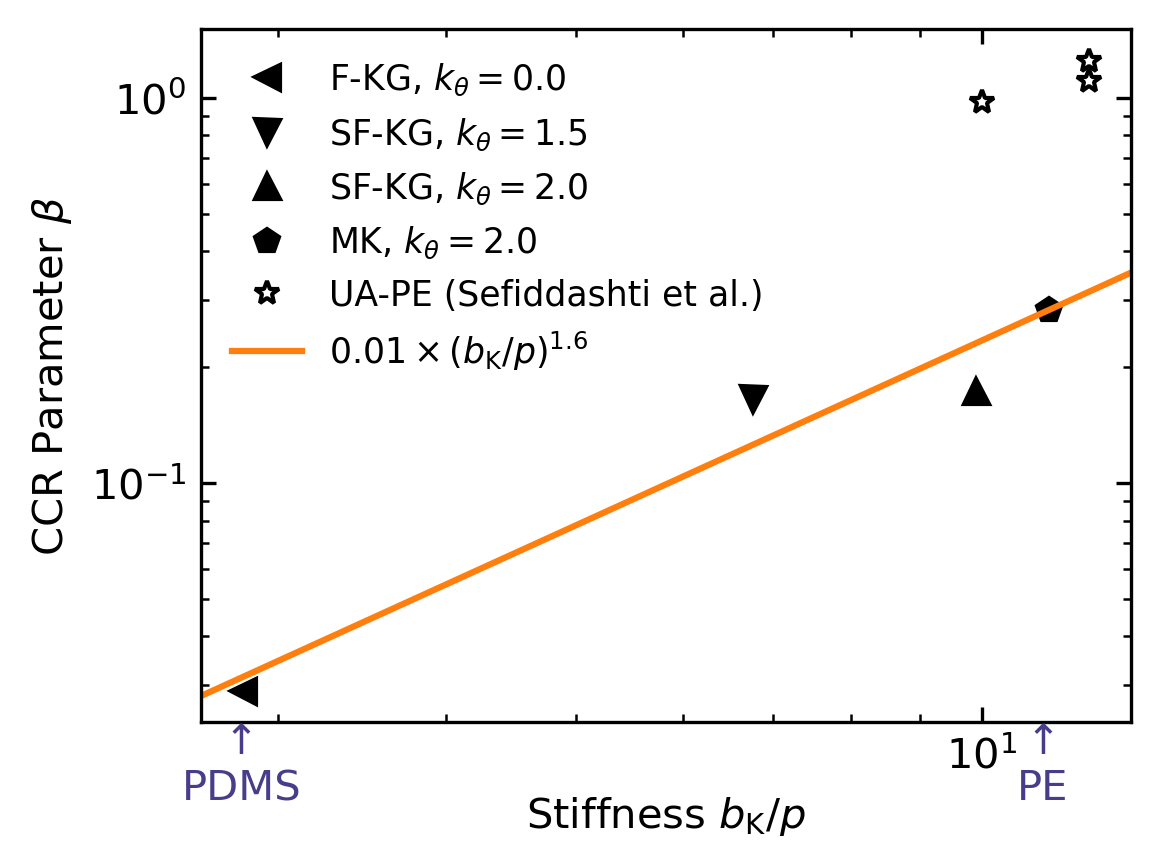}
    \caption{CCR parameter $\beta$ as a function of the Kuhn length to packing length ratio $b_{\textrm{K}}/p$ for our Kremer-Grest simulations (triangles), simulations by \citet{mk16model} (pentagons), and UA-PE simulations by \citet{sek-15, sek-16, sek-19, sek-19a} (stars), with a power law fit to the KG and MK data. Arrows on axis indicate stiffness of polyethylene (PE) and poly(dimethyl siloxane) (PDMS)~\cite{everaers20KG}}
    \label{fig:beta_nk}
\end{figure}

Fig.~\ref{fig:beta_nk} shows how the inferred value of $\beta$ depends on chain stiffness, quantified by the ratio of the Kuhn length $b_{\textrm{K}}$ to the packing length $p$.
\citet{dolata24chemical} suggested that $\beta$ should be a simple function of $b_{\textrm{K}}/p$, which they approximated by a power law $\beta \sim (b_{\text{K}}/p)^{1.9}$.
We find that a slightly less steep power law $\beta \propto (b_{\textrm{K}}/p)^{1.6}$ agrees with our KG simulations and the results from Mohagheghi and Khomami.
However, Dolata \textit{et al.} also computed $\beta$ and $b_{\text{K}}/p$ for a series of United-Atom Polyethylene (UA-PE) simulations performed by \citet{sek-15, sek-16, sek-19, sek-19a}, which do not agree with the computed power law.
These simulations are small enough that they likely suffer from the same self-entanglement issue as Dolata \textit{et al.}, inflating the measured value of $\beta$, but this effect may not be enough to account for the discrepancy.

\subsection{Observing Banding}

\begin{figure}[htb]
	\centering
	\includegraphics[width=\linewidth]{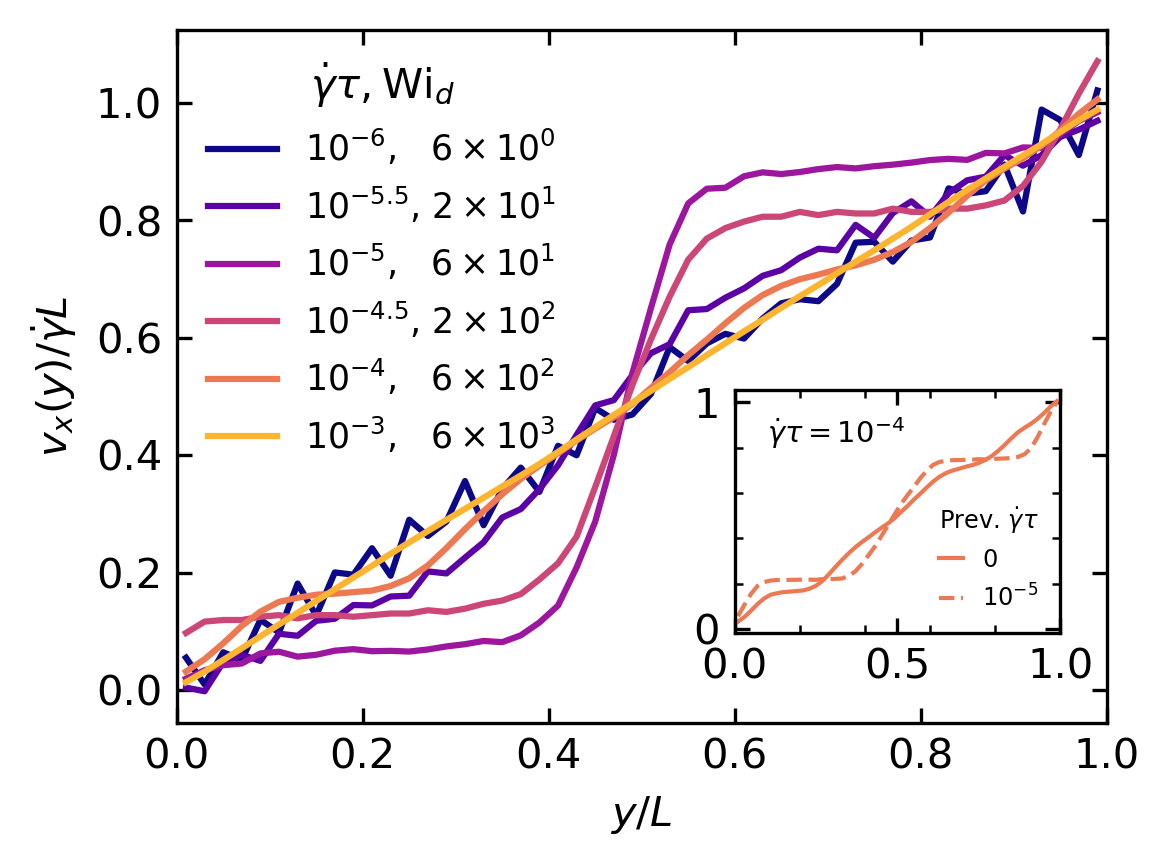}
	\caption{Steady-state velocity profiles of shear-banded Kremer-Grest simulations at various shear rates,  for a system with 3200 chains of length $N_{bead}=400$ with $k_\theta=2.0$. All profiles are  for systems starting from rest.
	The inset shows band profiles for two systems with different shear histories at the same shear rate $\gammadot\tau = 10^{-4}$: the solid profile is for a system starting from rest, as in the main figure, while the dashed profile is for a system initially sheared at $\gammadot\tau = 10^{-5}$ before increasing the shear rate to $\gammadot\tau = 10^{-4}$.}
	\label{fig:band_profiles}
\end{figure}

We observe steady-state shear banding in many of the shear simulations performed using the second shear procedure, depending on the length and stiffness of the chains and the rate at which they are sheared.
Fig.~\ref{fig:band_profiles} shows velocity profiles for a system with 3200 chains of 400 beads with $k_\theta=2.0$, corresponding to $\Zrhe=16$, at six different shear rates. All were started from an equilibrium
configuration and the shear velocities are averaged over 50 evenly-spaced bins in the gradient direction after the system reached steady-state.
At $\dot{\gamma}\tau = 10^{-3}$, the velocity profile is completely uniform, while at $10^{-4}$, it develops several small bands, and at $10^{-4.5}$ and $10^{-5}$ it clearly separates into one fast and one slow band.
At $10^{-5.5}$, the velocity gradient is small enough that local fluctuations become prominent, but there are still distinct bands at higher and lower shear rates, while at $10^{-6}$ only uncorrelated fluctuations remain, marking the end of shear banding.
These simulations lack the overall stress gradients present in experimental Taylor-Couette and cone-plate geometries, so configurations with multiple bands are more stable than in experiment.

\begin{figure}[htb]
    \centering
    \includegraphics[width=1\linewidth]{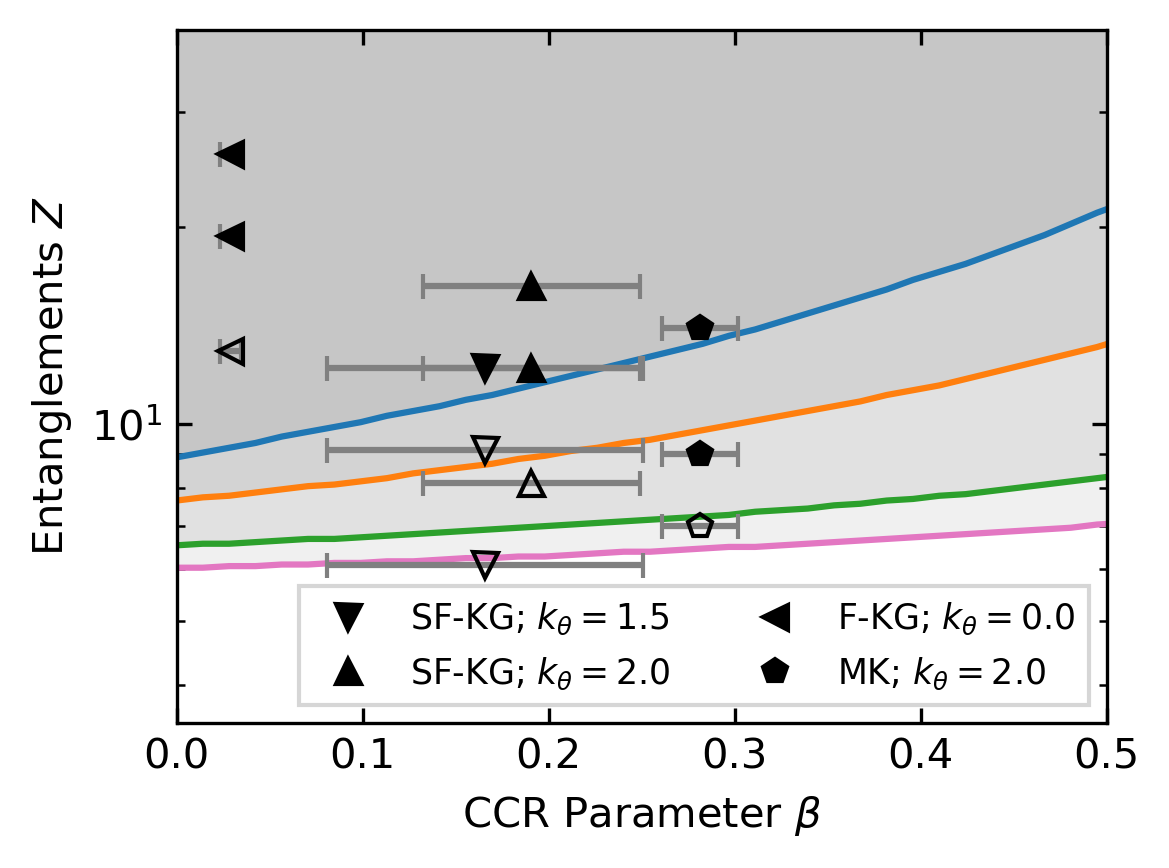}
	\caption{Material properties of simulations that do (solid $\blacktriangleleft,\blacktriangledown,\blacktriangle,\pentagofill$) and do not (hollow $\triangleleft,\triangledown,\triangle,\pentago$) band at some shear rates, overlaid on the predictions $\Zrhec(\beta,\alpha)$ of banding in Fig.~\ref{fig:phase_diagram}. Triangles $(\blacktriangleleft,\blacktriangledown,\blacktriangle)$ are flexible and semiflexible Kremer-Grest chains; pentagons $(\pentagofill)$ are from \citet{mk16banding}.}
    \label{fig:simulation_results}
\end{figure}

Other lengths and stiffnesses of chain follow the same pattern: flow is uniform at high shear rates, separates into bands at lower shear rates, and becomes uniform again at lower shear rates, for large $Z$.
This is in line with the behavior predicted by a non-monotonic constitutive curve, in which banding occurs at shear rates near the non-monotonic region of the constitutive curve.
Yet others do not band at any shear rate, as would be predicted by a monotonic constitutive curve.
Fig.~\ref{fig:simulation_results} shows which simulations exhibit banding, overlaid on the phase-space boundaries from Fig.~\ref{fig:phase_diagram}.
For each stiffness $k_{\theta}$ (and corresponding value of $\beta$), chains with $\Zrhe$ below some particular value $\Zrhec(k_\theta)$ never exhibit banding (hollow symbols), whereas those with $\Zrhe>\Zrhec(k_\theta)$ band at some shear rates (solid symbols).
In all cases, $\Zrhec(k_\theta)$ is close to that  predicted by the DO model for 
corresponding values of $\beta$ and 
plausible values of $\alpha$, despite the DO model being only a single-mode version of the more accurate microscopic GLaMM model \cite{glmm-03}.
However, our uncertainty in $\alpha$ limits the specificity of the DO model's predictions, and the most likely range of $\alpha$ indicated by literature data is higher than would best agree with $\Zrhec(k_\theta)$.
Furthermore, the simulations show a slight trend for $\Zrhec(k_\theta)$ to decrease for larger $\beta$, opposite the trend predicted by the model.
We can think of no mechanism which would produce such a trend, given the role of CCR in preventing banding.

\begin{figure}[htb]
    \centering
    \includegraphics[width=1\linewidth]{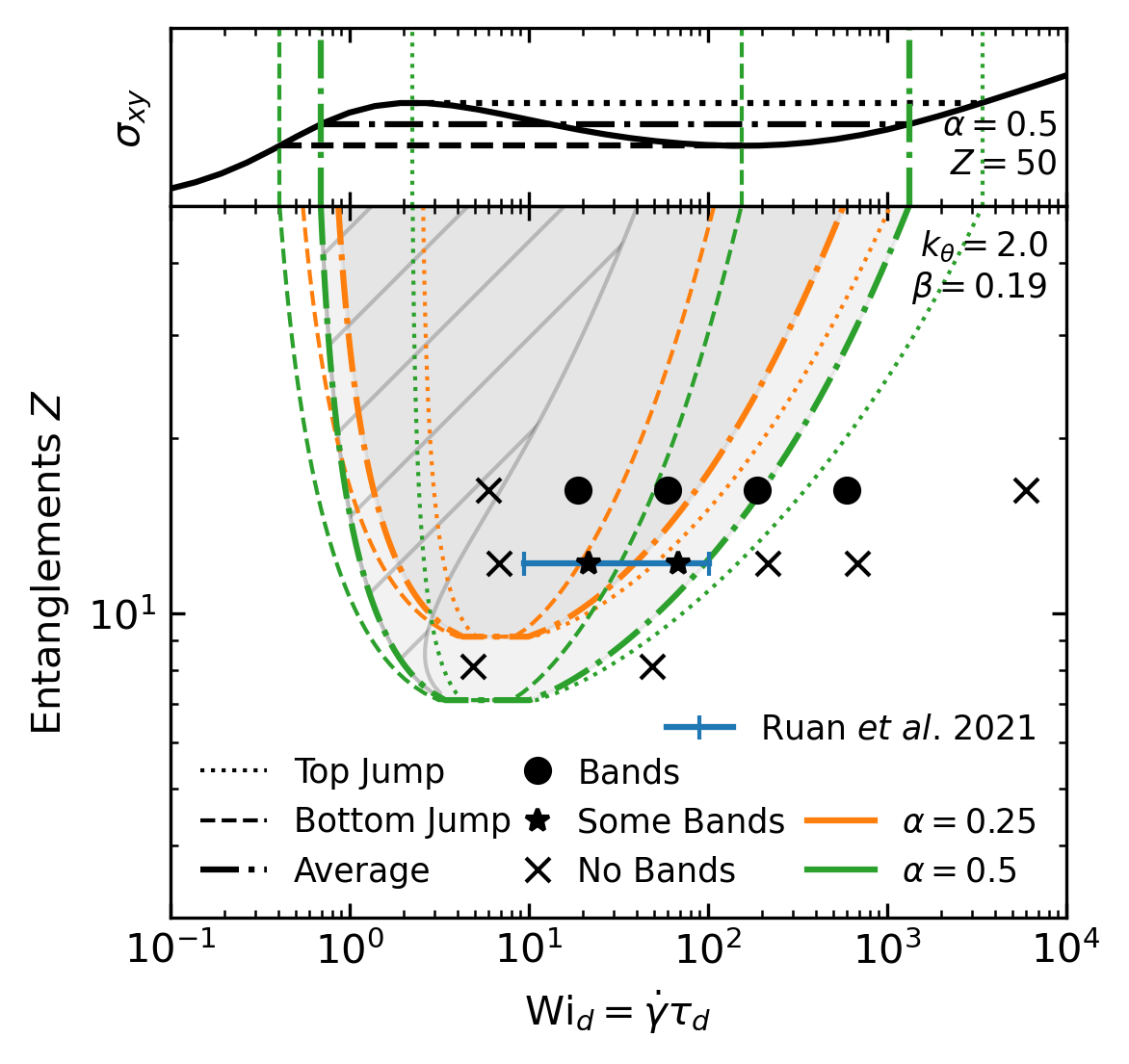}
    \caption{Range of reptation Weissenberg numbers for which the DO model predicts shear banding for a melt of chain lengths $\Zrhe$, with $\alpha=0.5$ or $0.25$, $\lmax\to\infty$, and $\beta(k_{\theta}\!\!=\!\!2)=0.21$ (fit in Table~\ref{table:sims}). Symbols represent SF-KG simulations with $k_\theta=2.0$ and whether they band in all, some, or no runs. The 
    blue line is the range of $\Wi_d$ where banding was measured by \citet{ruan_shear_2021}, for the same KG parameters. Line styles indicate method of calculating banding range, as illustrated in top panel and explained in text. Crosshatched region indicates where banding is expected in an infinite system but may be unobservable or absent due to finite-size effects.}
    \label{fig:plateau_width}
\end{figure}

Fig.~\ref{fig:plateau_width} shows the predicted reptation Weissenberg number $\Wi_d$ for banding, as a function of  $\Zrhe$, compared to the simulation data for $k_\theta=2.0$ with the fit value of $\beta = 0.19$; equivalent plots for other values of $k_\theta$ are in Appendix~\ref{app:range}.
The top row of points in this figure represent the system whose velocity profiles are shown in Fig~\ref{fig:band_profiles}, with uniform flow at the highest and lowest shear rates and banding in between.
Banding should occur at shear rates roughly in the shaded region of the plot, between the dash-dotted lines whose position depends on $\alpha$.
For any non-monotonic constitutive curve, such as that shown in Fig.~\ref{fig:plateau_width}a, uniform flow is strictly unstable in the range where $\sigma_{xy}$ decreases as a function of $\gammadot$, and should separate into bands in a process analogous to spinodal decomposition in fluid-fluid phase separation.
These bands must exhibit the same \textit{selected shear stress} $\sigma_{xy}$ to be stable, just as phase separated fluids must have the same pressure.
This selected stress could be determined by augmenting the DO model with gradient terms \cite{lu99,olmsted99a,olmsted99e} and solving for the stress at which the interface between shear bands is stationary, though we have not done so.

The selected stress plateau must be between the local maximum (``top jumping'') and local minimum (``bottom jumping'') on the constitutive curve, indicated by horizontal dotted and dashed lines in Fig.~\ref{fig:plateau_width}a.
We approximate the true selected stress as the average of these two values (the dash-dotted line).
Each of these potential plateau stresses intersects the stable parts of the constitutive curve at two points, indicated by vertical lines in Fig.~\ref{fig:plateau_width}a and plotted in Fig.~\ref{fig:plateau_width}b as a function of $\Zrhe$ for two different values of $\alpha$.
These two points are the predicted shear rates in the two bands, and in an infinitely-large system banding should occur exclusively at shear rates between them.

The predicted reptation Weissenberg number $\Wi_d$ that demarcate the banding region are consistently lower than the banding shear rates observed in the simulations, regardless of $\alpha$.
We have identified several factors that could contribute to this discrepancy: (1) high-shear-rate breakdown of tube models from instability and disentanglement, (2) history dependence of banding behavior, and (3) finite-size effects in our simulations.

The DO model and other tube models are expected to break down in the high-shear-rate limit, as the loss of the majority of the entanglements shown in Fig~\ref{fig:stress_strainrate} eliminates the constraints that define the tube, and as faster dynamical modes are activated that are not captured in these models.
Furthermore, \citet{LewyPRL2026} have shown that the RP model can develop 2D instabilities at these shear rates, which if present in the DO model would invalidate our assumption of locally-uniform shear.

Unbanded homogeneous flow has also been shown to be metastable in some conditions where banding is the most stable state \cite{olmsted99a,grand_slow_1997}, so it could require much longer simulations or different initial conditions to induce banding.
The presence or magnitude of banding can depend on the shear history of the system.
For instance, the inset of Fig.~\ref{fig:band_profiles} shows two systems which have reached steady state at the same final shear rate, but have different shear histories: the solid curve started directly from equilibrium, but the dashed curve was sheared at a rate with more pronounced bands first, resulting in more pronounced bands.
We also see the reverse effect, where initially shearing at a much higher (non-banding) shear rate can prevent banding at a rate where it otherwise occurs, consistent with metastability. For example, as shown in Fig~\ref{fig:band_history}a,
starting from a non-banded high shear rate of $\gammadot\tau=10^{-3}$ and reducing to $\gammadot\tau=10^{-5}$ (which forms bands when starting from rest), no shear band is observed, at least over the time scales accessible. The reverse process, in which we start at a lower shear rate and increase to a higher rate, generally gives the same result as starting from rest in terms of presence/absence of banding (as shown in Fig~\ref{fig:band_history}b of the appendix).
Continuously ramping the shear rate from rest to $\gammadot\tau=10^{-5}$ gives a similar long-time velocity profile to the abrupt startup.  
Similarly, the shear band disappears when increasing  from $\gammadot\tau=10^{-5}$, 
(which has a shear band) to $\gammadot\tau=10^{-3}$ (which does not).

Finite size effects can also prevent or obscure the formation of bands, particularly at low shear rates.
When computing the banding range predicted by the model from its constitutive curve, we assume that the fast and slow bands will have fixed shear rates $\gammadot_{\text{fast}}, \gammadot_{\text{slow}}$ at the selected stress plateau (on the constitutive curve), and that the bands will obey the lever rule \cite{olmstedbanding08}: they will resize themselves so that the overall shear rate $\gammadot_{\text{net}} = \phi_{\text{fast}}\gammadot_{\text{fast}} + \phi_{\text{slow}}\gammadot_{\text{slow}}$ is equal to the applied shear rate, where $\phi_{\text{fast}}=\ell_{\text{fast}}/L_y$ and $ \phi_{\text{slow}}=\ell_{\text{slow}}/L_y=1-\phi_{\text{fast}}$
are the thicknesses of the two bands as a fraction of the overall sample thickness $L_y$.
Expressed in terms of Weissenberg numbers, we thus have
\begin{equation}
 \phi_{\textrm{fast}}=
 \frac{\Wi_{d,\text{net}}-\Wi_{d,\text{slow}}}{\Wi_{d,\text{fast}}-\Wi_{d,\text{slow}}}
 \end{equation}

This assumption breaks down in systems with a finite thickness $L_y$ when the width of one of the bands 
becomes smaller than molecular sizes.
To estimate when this effect becomes relevant for simulations of various sizes, we assume that the high shear rate branch cannot be observed in practice when $\ell_{\text{fast}}$ drops below $\sim 3$ monomers in size or $\phi_{\textrm{fast}}<0.03$
(in our simulations, the box sizes (Table~\ref{table:simsmore}) are all of order $100\sigma$).
The crosshatched regions in Fig.~\ref{fig:plateau_width} (and Fig.~\ref{fig:plateau_width_alt} in Appendix~\ref{app:range}) correspond to shear rates in which $\phi_{\text{fast}}$ is between $0$ and $0.03$.
In principle, the region where $\phi_{\text{slow}}\equiv 1-\phi_{{\text{fast}}}$ is less than $0.03$ should also be shaded, but due to the logarithmic scale of the $\Wi_d$ axis, this region is negligibly small.

\section{Normal Stress Ratio}\label{sec:normal_stress}
In non-Newtonian fluids, the deviation of the stress from the isotropic pressure is frequently quantified in terms of \emph{normal stress differences} 
\begin{subequations}
    \begin{align}
        N_1 &= \sigma_{xx} - \sigma_{yy} \\
        N_2 &= \sigma_{yy} - \sigma_{zz}
    \end{align}
\end{subequations}
for stresses $\sigma_{xx}, \sigma_{yy}, \sigma_{zz}$ in the shear, gradient, and vorticity directions respectively.
Most complex fluids
reduce to Newtonian fluids in the linear regime at very low shear rates ($\Wi_d \ll 1$), so both normal stress differences scale quadratically with the shear rate $\gammadot$ as $\gammadot\to 0$ \cite{Bird}.
Consequently, the ratio $-N_2/N_1$ (which is positive for polymer fluids because $N_2$ is typically negative) is predicted to approach a constant as $\gammadot\to 0$.
The value of this constant is not clear; in the DE model, the limiting value is $1/7$ or $2/7$ depending on whether or not the Independent Alignment Approximation holds \cite{DoiEdwards_Book}; in the IM model, the limiting value is $1/4$ \cite{im_2001}; and in the DO model, it is $\alpha/2$ \cite{dolata22thermo}.
Identifying this limiting value from experiments or simulations would fix the value for $\alpha$ in our model, improving the precision of our predictions.

\begin{figure}[htb!]
\includegraphics[width=1\linewidth]{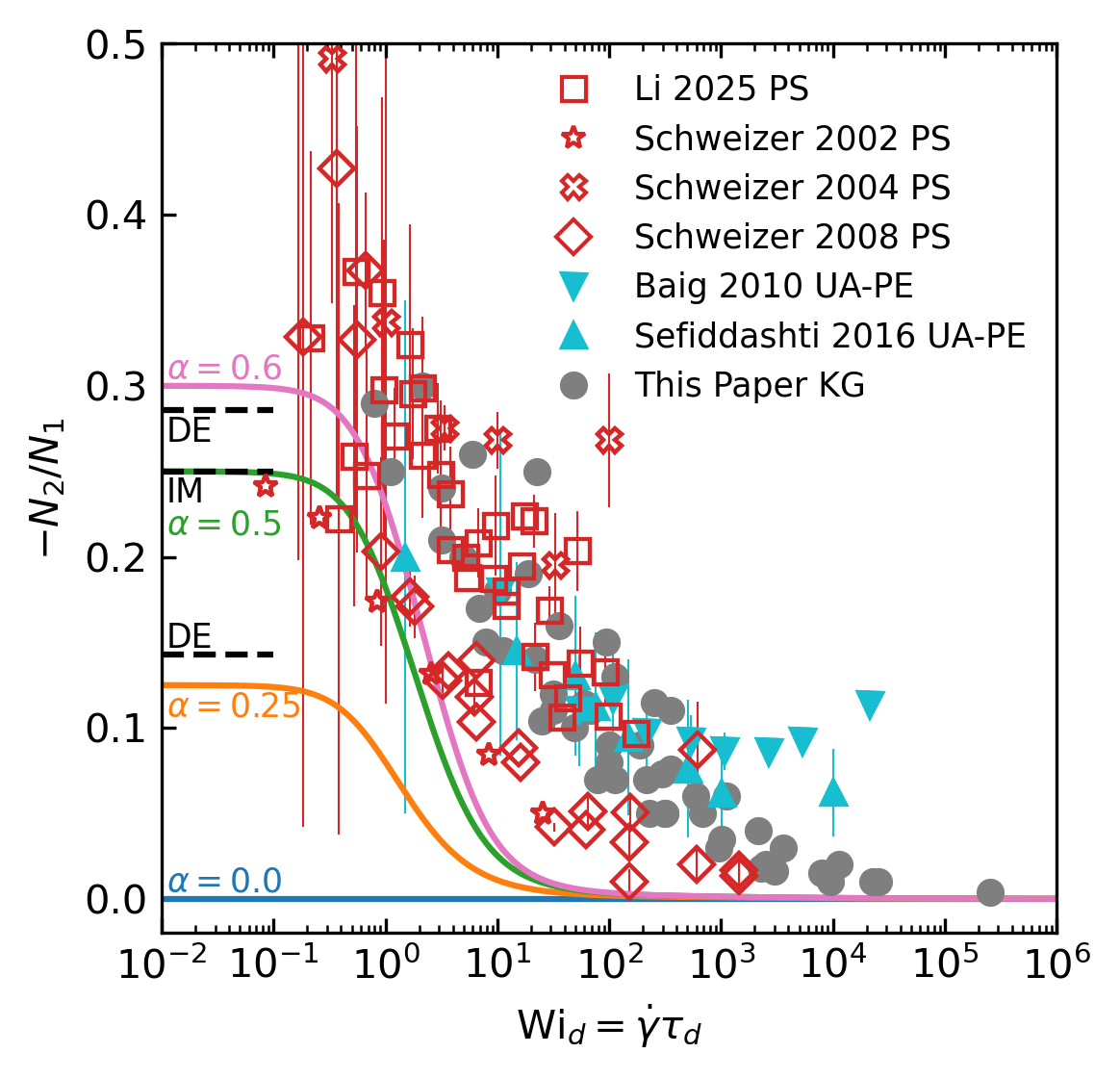}
    \caption{Normal stress ratios vs reptation Weissenberg numbers for various simulations (filled) and experiments (hollow) from the literature, along with predictions of the DO model with varying values of $\alpha$ (curves). Literature data from \citet{li_normal_2025}, \citet{schweizer2002measurement, schweizer2004nonlinear,schweizer_shear_2008}, \citet{bmk-10}, and \citet{sek-16}}
    \label{fig:exp_N2N1}
\end{figure}

Outside the linear regime most models predict that $N_2/N_1$ decreases with increasing shear rate $\Wi_d>1$, in the shear thinning regime.
Due to the quadratic dependence of both normal stress differences on the shear rate, the forces which must be measured to determine the normal stresses are orders of magnitude lower in the linear regime compared to the thinning regime, which makes it difficult to measure both regimes with the same apparatus.
Various groups \cite{keentok_coneplate_1982, ramachandran_dependence_1985, magda_concentrated_1994} have measured the normal stress ratio in well-entangled polymer solutions in the linear regime using flush-mounted pressure sensors and cone-partition-plate rheometers, and have found a nearly constant ratio of $-N_2/N_1 \approx 0.3$ (corresponding to $\alpha \approx 0.6$).
\citet{magda_concentrated_1994} also measured the normal stresses at shear rates above $1/\tau_d$ and found that $-N_2/N_1$ decreased as predicted, although instabilities such as edge fracture prevented their highest-concentration samples from reaching steady state in this regime.

Conversely, groups measuring the normal stress ratio in melts have mostly focused on the shear-thinning regime, despite the fact that polymer melts are even more susceptible to edge fracture and similar instabilities than solutions. 
Schweizer and co-workers\cite{schweizer2002measurement,schweizer2004nonlinear,schweizer_shear_2008,schweizer_cone-partitioned_2013}  and Vlassopoulos and co-workers \cite{costanzo_measuring_2018,li_normal_2025} measured the normal stress difference in several polystyrene melts, using cone-partitioned-plate rheometers to delay the onset of edge fracture.
Data from these groups is shown in Figure~\ref{fig:exp_N2N1}; due to the large uncertainty in the values of $N_2$ at low shear rates, the low-shear-rate limit cannot be determined with confidence, but the data suggest that $\alpha_{\textrm{melt}} \gtrsim \alpha_{\textrm{solution}}$.
To better compare data between different sources, we recalculated the reptation time for each of the melts in question using the entanglement Rouse time $\tau_e$ and the entanglement molecular weight $M_e$ derived by \citet{li_normal_2025}, as well as using their time-temperature superposition curve to estimate the temperature dependence of $\tau_e$, but applying the Likhtman relation (Eq.~\ref{eq:likhtman}) in the derivation rather than the simpler expression $\tau_d = 3 \Zrhe \tauR$ used in their calculations. 

Fig.~\ref{fig:exp_N2N1} shows all of the normal stress data from polymer melts and simulations of polymer melts that we could find in the literature.
More information on these simulations is Table~\ref{table:N2N1}.
Most of the data, both experimental and simulation, is for large Weissenberg numbers, and none of the data show the plateau expected at low shear rates where both $N_1,N_2\sim\gammadot^2.$

Measurements of the normal stress ratio in simulations are similarly focused on the shear-thinning regime.
Low-shear-rate simulations require much more simulation time to reach steady state than high-shear-rate ones, and measurements of $N_1$ and $N_2$ are much more susceptible to fluctuation noise due to their smaller magnitude, so the linear regime is impractical to reach with current methods and hardware. 
Outside of these restrictions the normal stress differences are much more straightforward to measure in simulation than in experiment, and edge fracture is  eliminated due to periodic boundary conditions.
\citet{bmk-10} and \citet{sek-16} measured the normal stress ratio in their simulations of UA-PE (which has dihedral angles and rigid bond lengths and angles), which we reproduce in Fig.~\ref{fig:exp_N2N1}, along with measurements from our own KG simulations (which has only springs and no prescribed bond or dihedral angles).

The resulting $-N_2/N_1$ vs $\Wi_d$ curves collapse reasonably well between experiment and the two simulation types, in line with the DO model's prediction that the normal stress ratio primarily only depends on $\Wi_d$ and $\alpha$.
This suggests that the value of $\alpha$ may be universal between different polymer melts, but the data at low $\Wi_d$ is insufficient to determine its value.
The decay in $-N_2/N_1$ occurs over several orders of magnitude of $\Wi_d$, whereas the DO model predicts a more rapid drop off (solid lines); more detailed models may match this decay better.

The UA-PE simulations diverge somewhat from the KG data at high shear rates, which we speculate is due to sub-entanglement-scale differences between the two polymer representations (UA-PE vs KG); the divergence begins when $\Wi_d \sim 10^3$, at which point $\gammadot\sim \tau_e^{-1}$.

\section{Conclusions}\label{sec:conclusion}
The DO model makes specific predictions for the conditions under which entangled polymer melts and solutions should exhibit shear banding.
These predictions agree qualitatively and semi-quantitatively with molecular dynamics simulations of bead-spring polymers.
Shear banding is controlled by $\Zrhe$, the normal stress ratio $-\alpha/2$, and the CCR parameter $\beta$. The normal stress ratio is notoriously difficult to determine for polymer melts, although its value appears to be universal.
Regardless of the value of $\alpha$, for the small values of $\beta\simeq0.03\text{--}0.3$ that best fit the simulations the DO model predicts that shear banding should occur for chains with $\Zrhe \gtrsim 10$, which agrees with simulations.
The overestimate by the model of the range of shear rates that support banding suggests that the model is inadequate at high shear rates for significant degree of distentanglement, as noted by \citet{wangview2008a}.

The prediction that the entanglement cutoff for banding $\Zrhec$ should increase with increasing $\beta$ and thus increasing $k_\theta$ does not appear to hold for the range of stiffnesses in our simulations.
Given the role that convective constraint release plays in avoiding nonmonotonic constitutive curves, it is unexpected that more rapid disentanglement should be correlated with a higher tendency to shear band rather than a lower one.
There may be some confounding effect that contributes to this discrepancy, such as the qualitative difference in the bending potentials of flexible and semiflexible KG chains.
However, there may also be a flaw in the methodology of measuring $\beta$: the model's description of high-shear-rate entanglement dynamics may be inadequate, or the identification of kinks with entanglements may break down under flow or alignment.

A similar polymer-specific variation of constraint release was found by \citet{watanabe_nonuniv_2024}, who measured \emph{equilibrium} constraint release experimentally.
In contrast, they found that polymers with stronger equilibrium constraint release tended to be \emph{less} stiff, which if applicable to CCR would be consistent with the trend in Fig.~\ref{fig:simulation_results}.
However, it is also possible that the detailed microscopic mechanisms of equilibrium constraint release and convected constraint release (termed `entanglement stripping' by \citet{hhhr-15}) differ significantly.

Analysis \cite{dolata24chemical} of data from UA-PE models \cite{sek-15,sek-16,sek-19,sek-19a} for polyethylene shows a substantially larger $\beta\simeq 1$.
This estimate of $\beta$ is likely inflated due to finite system size leading to self-entanglement, but these effects may not be sufficient to account for the discrepancy from bead-spring simulations.
United-atom polyethylene has a single dihedral angle for every interior atomistic monomer on the chain, compared to the three degrees of freedom for each (coarse-grained) `monomer' in the KG bead-spring model.
This structural difference may account for the increased strength of convective constraint release above that predicted by the chain stiffness, and the richer microscopic detail of fully-atomistic polymers with bending and stretching energies could support similar large values for $\beta$.
For these values, the predicted crossover value for banding $\Zrhec(\beta,\alpha)$ is much larger and more sensitive to $\alpha$, perhaps as high as $\Zrhec=20\textrm{--}300$ depending on $\alpha$, which may contribute to the lack of banding observed in experimental melts of comparable stiffness.
If the trend in $\beta$ vs. $b_k/p$ shown in Fig~\ref{fig:beta_nk} holds for experiments, and if the predicted trend that higher $\beta$ leads to higher $\Zrhec$ is accurate for larger differences in $\beta$, we expect that flexible polymers such as PDMS (for which $b_{\textrm{K}}/p=2.8$~\cite{everaers20KG}) would be more likely to exhibit banding.

\section*{Acknowledgments}
We thank Joe Peterson, Ben Dolata, Thomas O'Connor, Marco Galvani Cunha, and Shi-Qing Wang for discussions; and Martin Kr\"oger for generous help with and advice about the Z1+ code. PDO and LLN thank the National Science Foundation (DMREF-2118769), and PDO thanks Georgetown University and the Ives Foundation for financial support. This work was performed, in part, at the Center for Integrated Nanotechnologies, an Office of Science user facility operated by the U.S. Department of Energy (DOE) Office of Science. Sandia National Laboratories is a multi-mission laboratory managed and operated by National Technology and Engineering Solutions of Sandia, LLC, a wholly owned subsidiary of Honeywell International, Inc., for the U.S. DOE National Nuclear Security of Administration under contract no. DENA-0003525. The views expressed in this article do not necessarily represent the views of the U.S. DOE or the United States government.

\section*{Author Declarations}
The authors have no conflicts to disclose.

\section*{Data Availability}
The data that support the findings of this study are available from the corresponding author upon reasonable request.


\appendix

\section{Simulation Details}
\label{app:simsmore}

Table~\ref{table:simsmore} lists the specific systems sizes of all the simulations we have performed, along with the three relevant timescales $\tau_e, \tauR, \tau_d$ of each system and which of the shear procedures has been used for that system.
For systems with non-cubic simulation boxes, $L_{y,z}$ is the box length in the gradient and vorticity directions and $L_x$ is the length in the shear direction, so $L_x/L_{y,z}$ is the aspect ratio of the box.

\begin{table}[b]
\begin{tabular}{lllllllll}
\hline \hline\\[-11truept]
$k_\theta$ & $N_{\text{mon}}$ & $N_{\text{chains}}$ & $\dfrac{L_{y,z}}{\sigma}$ & $\dfrac{L_{x}}{L_{y,z}}$ & $\dfrac{\tau_e}{\tau}$   & $\dfrac{\tau_R}{\tau}$   & $\dfrac{\tau_d}{\tau}$   & Type    \\
           &                  &                     & $\times10^{-1}$  &                 & $\times10^{-3}$ & $\times10^{-5}$ & $\times10^{-6}$ &         \\
\hline
0          & 500              & 500                 & 6.7              & 1               & 10.9            & 4.5             & 1.9             & Neither \\
0          & 500              & 16000               & 13.3             & 4               & 10.9            & 4.5             & 1.9             & SLLOD   \\
0          & 1000             & 2000                & 13.3             & 1               & 10.9            & 18.1            & 24.5            & Both    \\
0          & 1500             & 1000                & 12.1             & 1               & 10.9            & 40.8            & 101.7           & DPD     \\
0          & 2000             & 1000                & 13.3             & 1               & 10.9            & 72.5            & 272.3           & DPD     \\
1.5        & 200              & 250                 & 3.9              & 1               & 1.7             & 0.6             & 0.2             & SLLOD   \\
1.5        & 200              & 2000                & 7.8              & 1               & 1.7             & 0.6             & 0.2             & DPD     \\
1.5        & 300              & 3200                & 10.4             & 1               & 1.7             & 1.4             & 1.1             & DPD     \\
1.5        & 400              & 400                 & 5.7              & 1               & 1.7             & 2.6             & 3.2             & SLLOD   \\
1.5        & 400              & 1600                & 9.1              & 1               & 1.7             & 2.6             & 3.2             & DPD     \\
1.5        & 400              & 3200                & 11.5             & 1               & 1.7             & 2.6             & 3.2             & DPD     \\
1.5        & 400              & 12800               & 11.5             & 4               & 1.7             & 2.6             & 3.2             & SLLOD   \\
1.5        & 800              & 400                 & 7.2              & 1               & 1.7             & 10.3            & 35.6            & SLLOD   \\
2          & 200              & 2000                & 7.8              & 1               & 1.2             & 0.8             & 0.5             & DPD     \\
2          & 300              & 400                 & 5.2              & 1               & 1.2             & 1.7             & 2.2             & Both    \\
2          & 300              & 3200                & 10.4             & 1               & 1.2             & 1.7             & 2.2             & Both    \\
2          & 300              & 10800               & 15.6             & 1               & 1.2             & 1.7             & 2.2             & DPD     \\
2          & 300              & 12800               & 10.8             & 4               & 1.2             & 1.7             & 2.2             & SLLOD   \\
2          & 400              & 3200                & 11.5             & 1               & 1.2             & 3.1             & 6.0             & DPD     \\
2          & 400              & 12800               & 11.5             & 4               & 1.2             & 3.1             & 6.0             & SLLOD  
\\ \hline \hline
\end{tabular}
\caption{Numbers of chains, box sizes $L_x$ and aspect ratios $L_x/L_{y,z}$, timescales $\tau_e,\tauR,\tau_d$, and shear algorithms used in all KG simulations performed. The simulation with neither shear type was used only as an equilibrium configuration.}
\label{table:simsmore}
\end{table}

\section{Additional Banding Range Plots} \label{app:range}

\begin{figure}[htb!]
    \centering
    \includegraphics[width=1\linewidth]{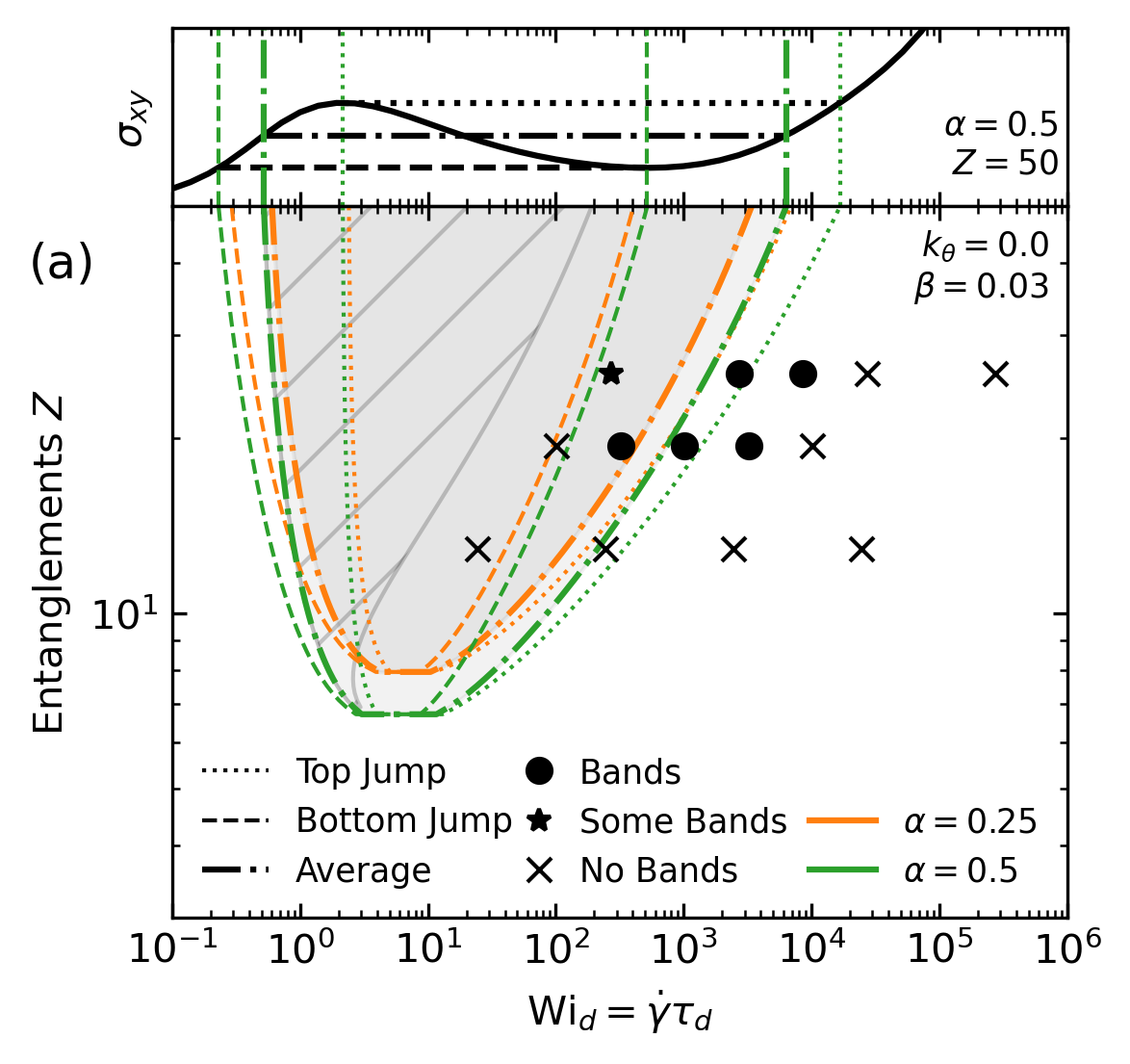}
    \includegraphics[width=1\linewidth]{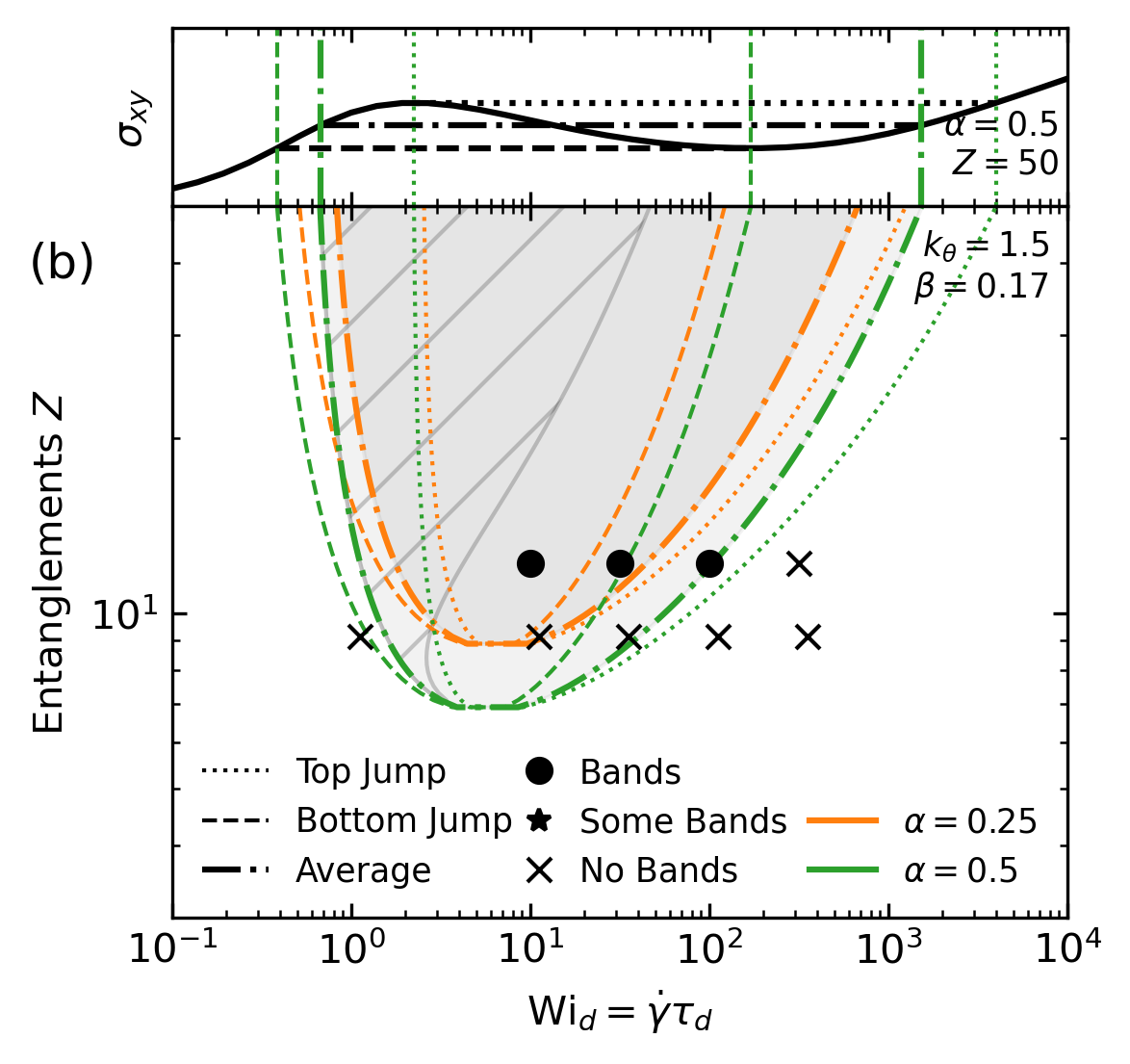}
    \caption{Range of reptation Weissenberg numbers for which the melt DO model predicts shear banding to occur for different chain lengths $\Zrhe$, with (a) $\beta\simeq0.03\,  (k_{\theta}=0.0)$ and (b)  $\beta\simeq0.17\,  (k_{\theta}=1.5)$.}
    \label{fig:plateau_width_alt}
\end{figure}

Fig.~\ref{fig:plateau_width} of the main text compares the range of banding shear rates between theory and simulation for chains with bending stiffness $k_\theta=2.0$; Fig.~\ref{fig:plateau_width_alt} shows the equivalent data for $k_\theta=0$ and $k_\theta=1.5$.
The simulations with $k_\theta=1.5$ agree with the theory better than $k_\theta=2.0$, but those with $k_\theta=0$ agree considerably worse, in line with their greater discrepancy in $\Zrhe_c$.

\section{Effects of $\varepsilon$ and $\lmax$}
\label{app:eps_lmax}
\begin{figure}[htb!]
	\centering
	\includegraphics[width=1\linewidth]{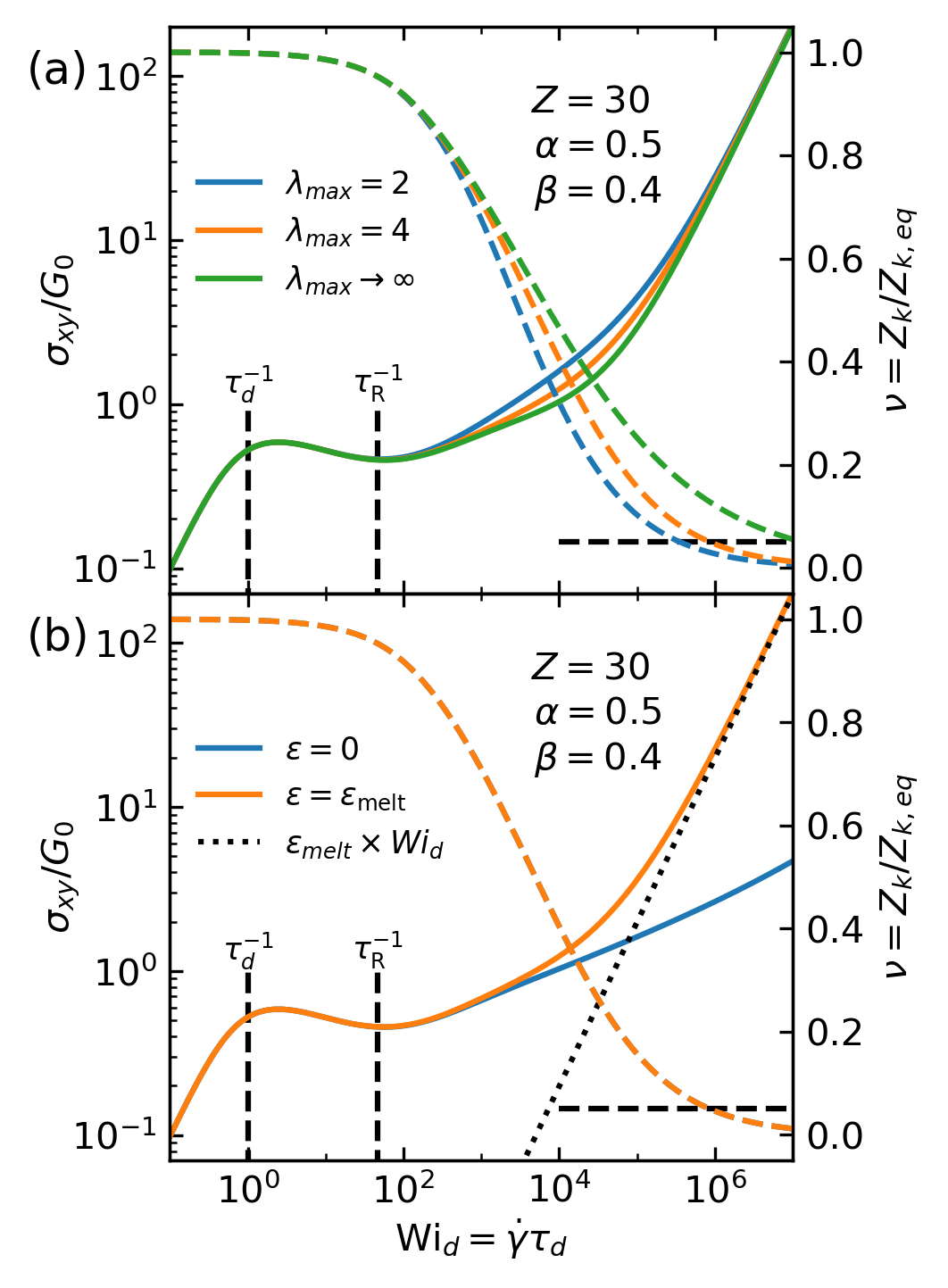}
	\caption{Effect of varying $\lmax$ (a) and $\varepsilon$ (b) on constitutive curves (solid) and disentanglement (dashed) for the DO model, as in Fig.~\ref{fig:stress_strainrate}.  The black dashed line in each figure corresponds to $Z_k=3$  kinks in the tube, where the tube model probably fails; while the black dotted line in (b) is the stress from the background viscosity}
	\label{fig:stress_minor}
\end{figure}

Fig.~\ref{fig:stress_minor} illustrates the effects of varying $\lmax$ and $\varepsilon$ on the constitutive curves of the DO model. Neither parameter has much effect on monotonicity, but they do affect the constitutive curve at relatively high shear rates, which are typically outside the non-monotonic regime.

The maximum stretching ratio $\lmax$ has a weak effect because the departure from linear elastic behavior only becomes relevant to steady-state banding once the tube has stretched considerably, which only happens for shear rates exceeding $1/\tauR$.
For smaller values of $\Zrhe$, $1/\tauR$ is closer to the start of the plateau regime, so the slight increase in stress due to smaller $\lmax$ is more likely to eliminate a decreasing regime.
The value of $\lmax$ does have a substantial impact on tube disentanglement at high shear rates, so it is always relevant for fitting $\beta$ from disentanglement data in simulations.

\begin{figure}[htb!]
    \centering
    \includegraphics[width=0.9\linewidth]{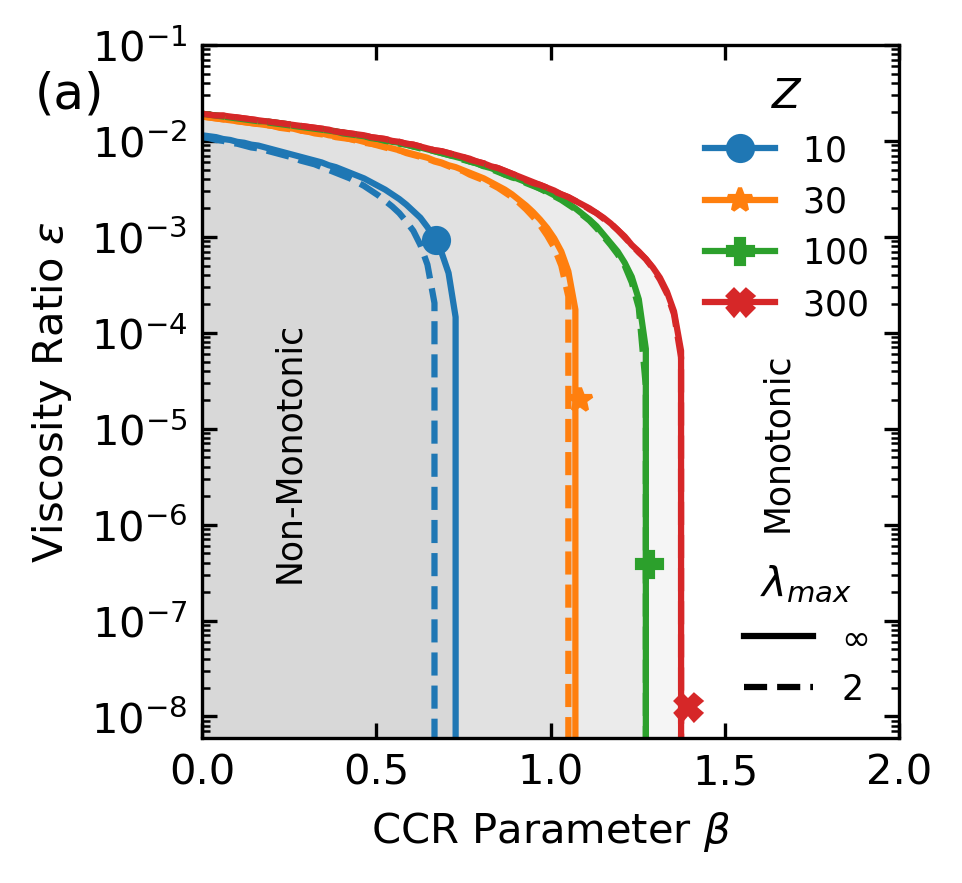}
    \includegraphics[width=0.9\linewidth]{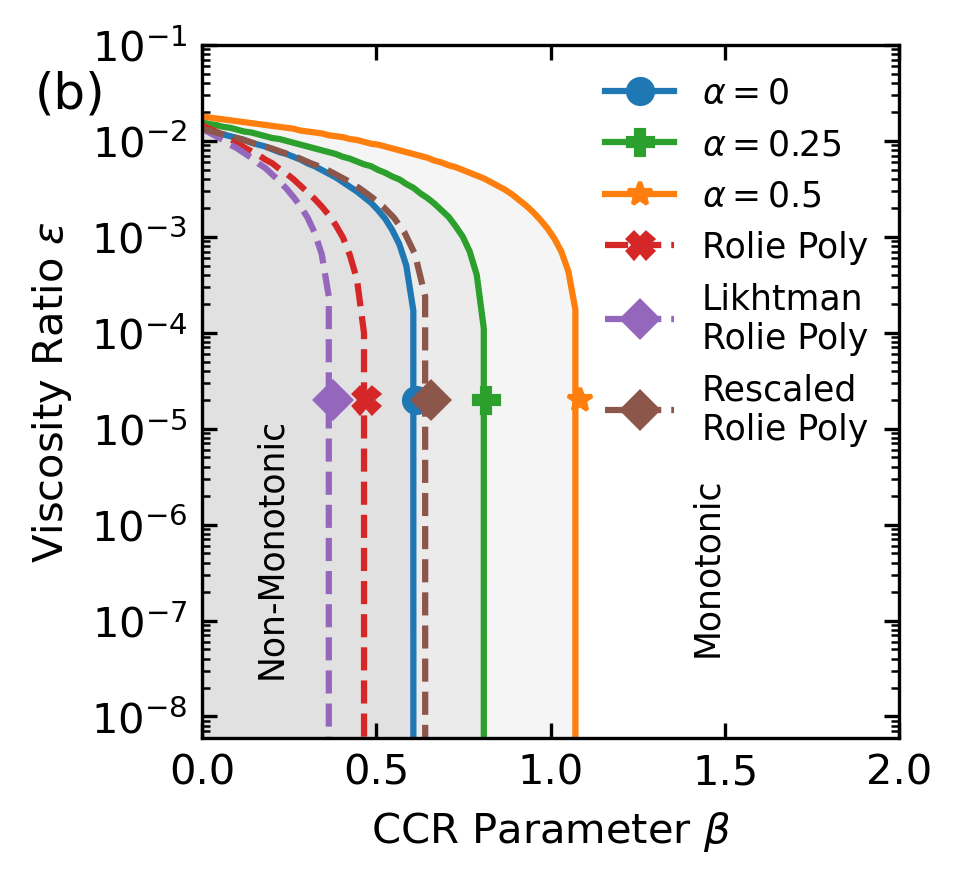}
    \caption{
	Phase-space boundaries identifying non-monotonicity in $\beta,\varepsilon$ space, either (a) varying $\Zrhe$ and $\lmax$ while keeping $\alpha=0.5$, or (b) varying $\alpha$ while keeping $\Zrhe=30, \lmax\to\infty$.
	The markers on the curves are placed at the value $\varepsilon = \varepsilon_{\textrm{melt}}(\Zrhe)$.
	(b) includes boundaries for the original  Rolie-Poly model with $\tau_d=3\Zrhe\tau_{\textrm{R}}$  \cite{likhtmangraham03};
     an updated version using the Likhtman relation (Eq.~\ref{eq:likhtman}) for $\tau_d$; and a version in which, in addition, the value of $\beta$ has been rescaled according to Eq.~\ref{eq:nonstretch}.}
    \label{fig:eps_stability_old}
\end{figure}

The viscosity ratio $\varepsilon$ becomes relevant only when the associated stress (black dotted line in Fig.~\ref{fig:stress_minor}) becomes significant relative to the tube stress.
While this can occur in the plateau regime and affect monotonicity, it only does so for relatively large $\varepsilon$ and, as $\varepsilon$ scales as $\Zrhe^{-3}$, is only relevant for small $\Zrhe$.
Fig.~\ref{fig:eps_stability_old}a illustrates this: the boundaries between monotonic and non-monotonic constitutive curves in $\varepsilon\textrm{-}\beta$ space become independent of $\varepsilon$ for $\varepsilon \lesssim 10^{-4}$. The values of $\varepsilon_{\textrm{melt}}$, illustrated by the symbols along the boundaries, are well below that point for all $\Zrhe \gtrsim 10$.

Fig.~\ref{fig:eps_stability_old} shows several alternate slices through the monotonicity phase space of the model, illustrating the effect of $\varepsilon$ more clearly.
Figure \ref{fig:eps_stability_old}a is analogous to Fig.~3 of \citet{adams11RP}, with a logarithmic scale for $\varepsilon$.
Fig.~\ref{fig:eps_stability_old}b illustrates the effects of varying $\alpha$ in this perspective, but also allows for direct comparison to several versions of the Rolie-Poly model, discussed below.

\section{Rolie-Poly Comparison}

\begin{figure}[htb!]
    \centering
    \includegraphics[width=1\linewidth]{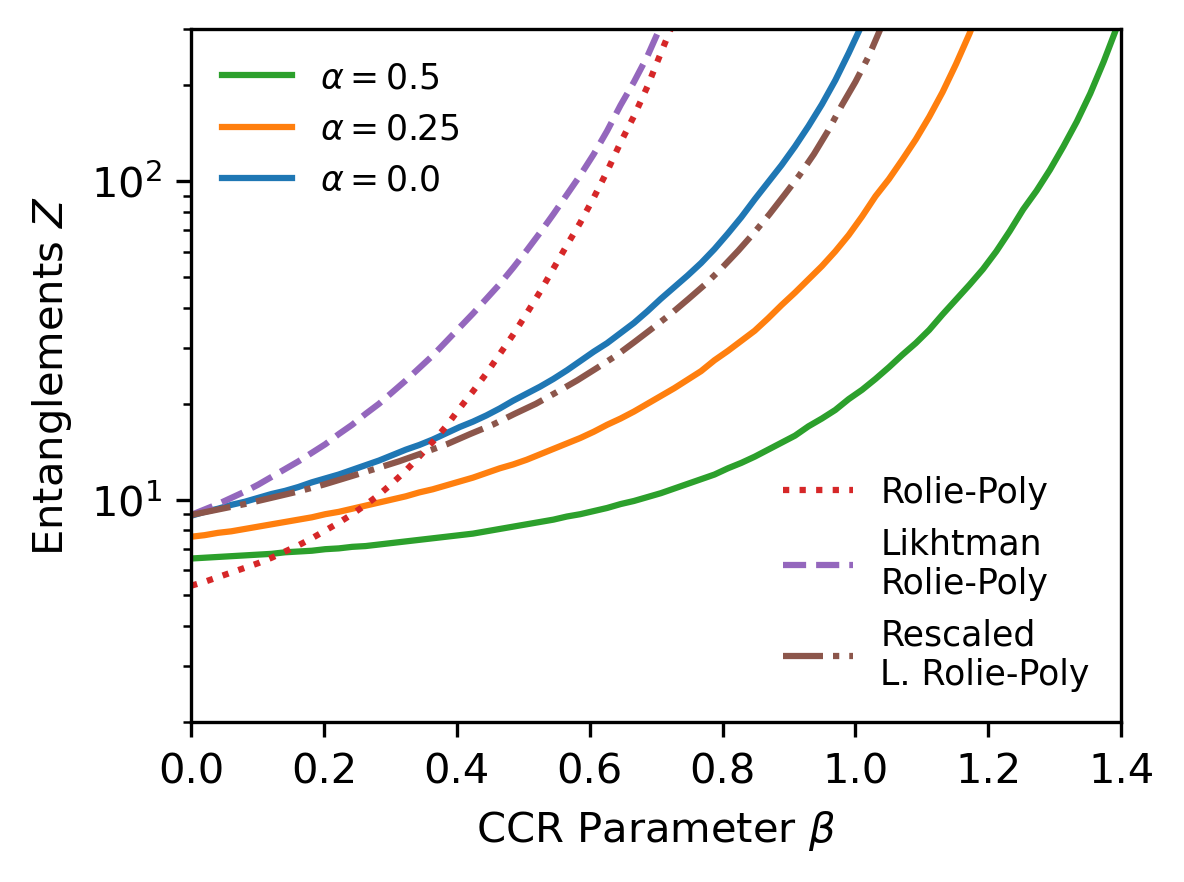}
    \caption{Monotonicity boundaries in $\Zrhe, \beta$ space for varying $\alpha$ (with $\lmax\to\infty$ and $\varepsilon=\varepsilon_{\text{melt}}$), as well as for three variants of the Rolie-Poly model (for which $\alpha=0$).}
    \label{fig:RP_boundary}
\end{figure}

The DO model is directly based on the Rolie-Poly model, with $\Zrhe$ and $\beta$ corresponding directly to equivalently-named parameters in that model.
It incorporates several additional features, such as a dynamical equation of motion for  disentanglement, finite extensibility, anisotropic relaxation, and conformation-dependent relaxation;  if we eliminate some of these features by setting $\alpha=0$ and $\lmax\to\infty$, it should be quite similar to the RP model.
Fig.~\ref{fig:RP_boundary} shows how phase boundaries in Fig.~\ref{fig:phase_diagram} compare to those predicted by the Rolie-Poly model (dotted), along with two variations of the RP model.

The first variation (dashed) is the `Likhtman' Rolie-Poly model, in which the relationship $\tau_d = 3\Zrhe\tauR$ in the original model has been replaced with the more detailed Likhtman relation $\tau_d = 3 \Zrhe f_L(\Zrhe)\tauR$ derived by \citet{lm02}, which is used in the DO model.
The difference between the two becomes negligible in the high-$\Zrhe$ limit, but applying the correction makes the two models identical when $\beta=0$.

The second variation is due to \citet{chen25critical}, who investigated both models in the high-$\Zrhe$, nonstretching limit. By expanding both models in terms of $1/\Zrhe$, they found that the two models become identical in this limit when $\alpha=0$, except for a rescaling in the value of $\beta$:
\begin{equation} \label{eq:nonstretch}
	\beta_{RP} = \tfrac{1}{2} \left( 1 + \tfrac{1}{3} \beta_{DO}^2 \right)\beta_{DO}
\end{equation}
(This is a corrected version of Eq.~(50) of \cite{chen25critical}; per private communication, the term in parentheses was incorrectly given as $1-\tfrac{1}{3}\beta_{DO}^2$).
By replacing $\beta$ in the RP model with this rescaled value, in addition to applying the Likhtman correction factor, we get a Rescaled Likhtman Rolie-Poly model, the dash-dotted line in Fig.~\ref{fig:RP_boundary}, whose monotonicity behavior is close to the DO model with $\alpha=0$.

\section{Banding profiles}

\begin{figure}[htb]
	\centering
	\includegraphics[width=1\linewidth,trim={0 8cm 0 0},clip]{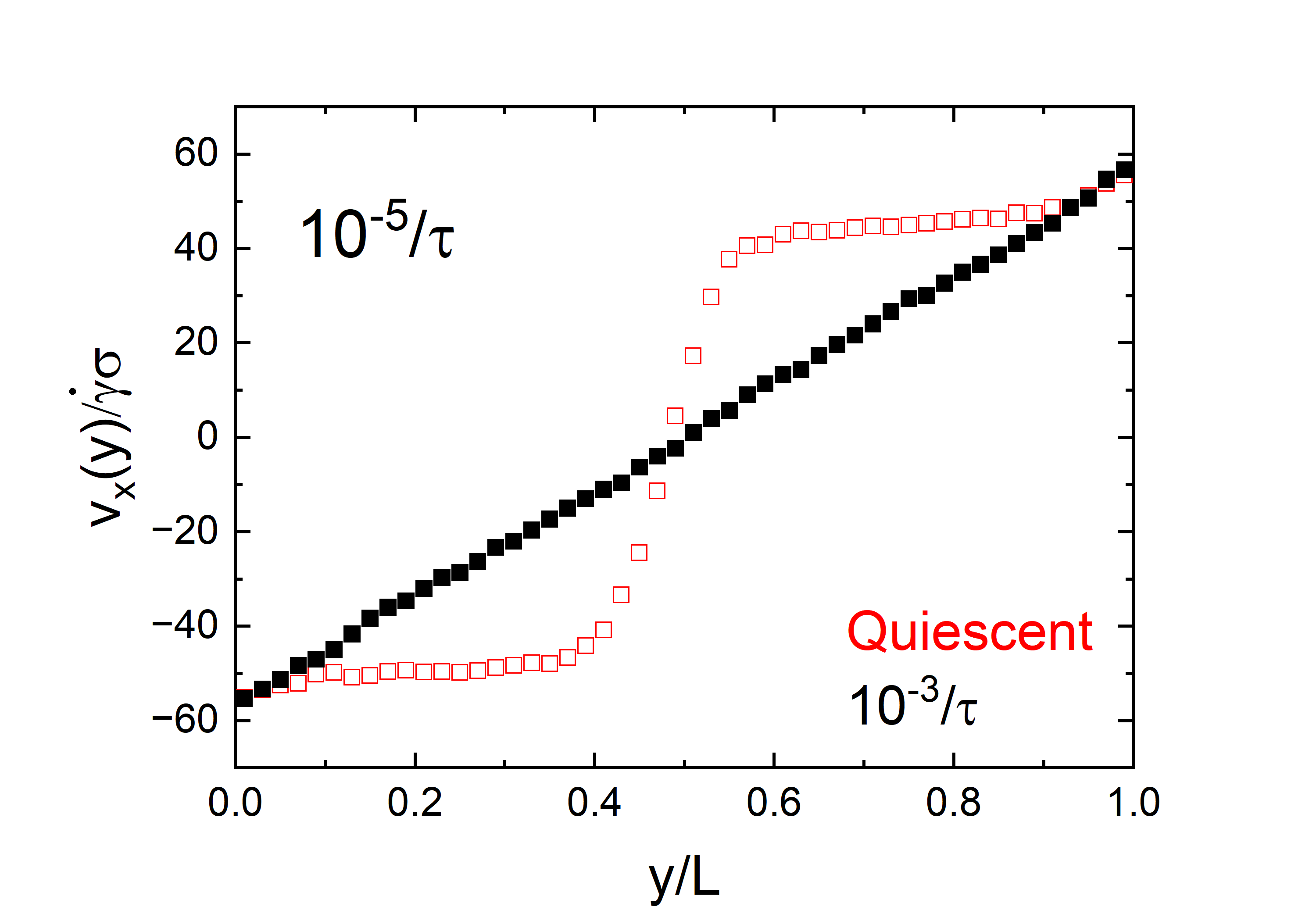}
	\includegraphics[width=1\linewidth,trim={0 0 0 5cm},clip]{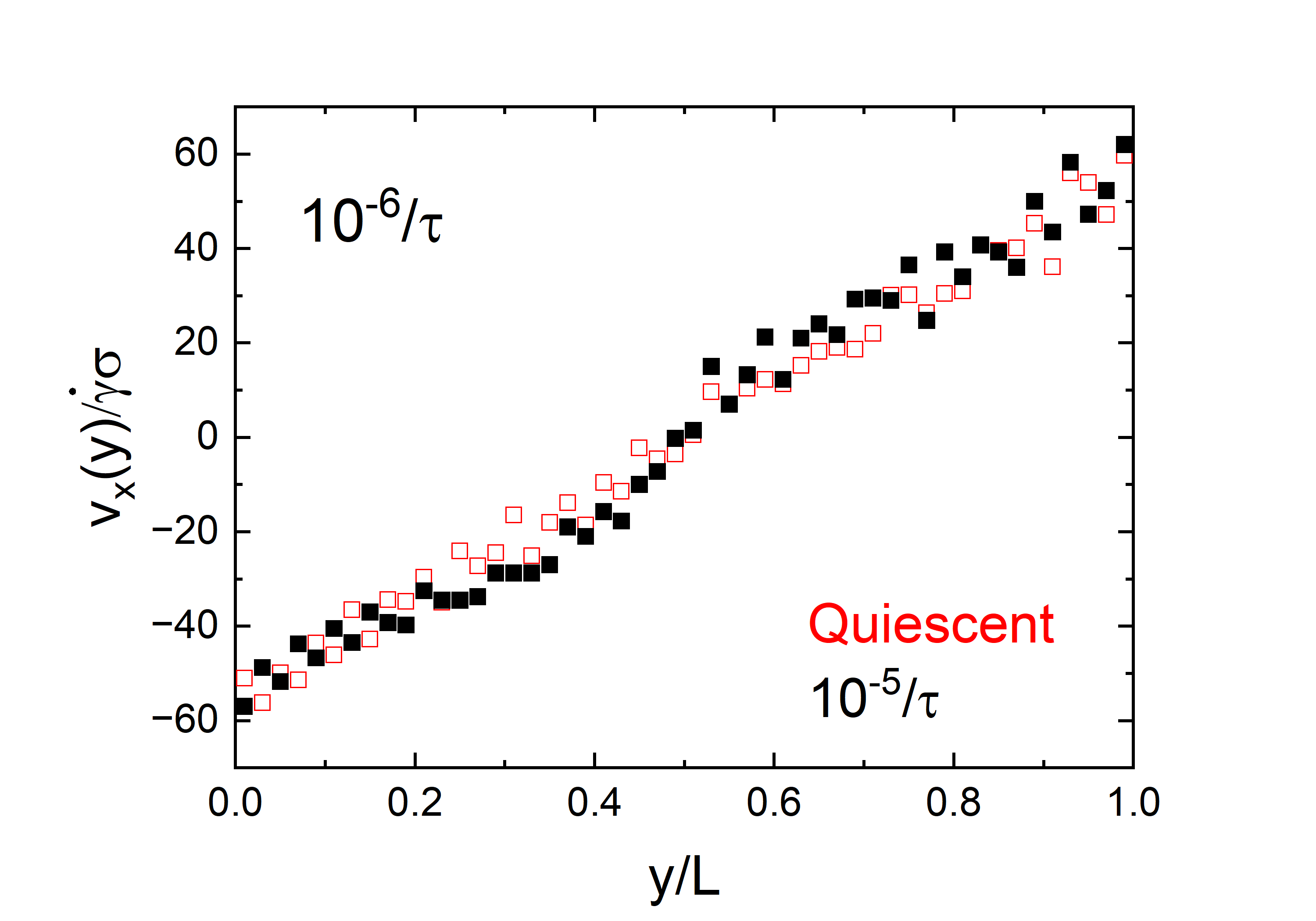}
	\caption{Steady-state velocity profiles with different shear histories for a system of 3200 chains of 400 beads with $k_\theta=2.0$. In each part, two steady-state profiles are shown for the same shear rate ($\gamma = 10^{-5}$ and $10^{-6} \tau^{-1}$ starting either from rest (red, hollow symbols) or from a steady-state configuration at a different shear rates $(10^{-3}$ and $10^{-5} \tau^{-1}$) (black, solid symbols).}
	\label{fig:band_history}
\end{figure}

Figure~\ref{fig:band_history} presents steady state profiles for a system of 3200 chains of 400 beads with $k_\theta=2.0 $ for different shear histories

\section{Normal Stress Ratio Data}
Table~\ref{table:N2N1} lists some properties of the polymers in the previous experiments and simulations measuring the normal stress ratio which are plotted in Figure~\ref{fig:exp_N2N1}.

\begin{table}[]
\begin{tabular}{ccccp{1cm}c}
\hline\hline
Source                          & Polymer & Length     & $M_w/M_n$ & $T$  & $Z$    \\ \hline
\cite{li_normal_2025}           & PS      & 283~kDa    & 1.01      & 160\degree~\!\!C, 170\degree~\!\!C & 16.6 \\
\cite{li_normal_2025}           & PS      & 143~kDa    & 1.11      & 130\degree~\!\!C, 160\degree~\!\!C & 8.4  \\
\cite{schweizer2002measurement} & PS      & 158~kDa    & 2.85      & 190\degree~\!\!C & 9.3  \\
\cite{schweizer2004nonlinear}   & PS      & 200~kDa    & 1.06      & 175\degree~\!\!C & 11.8 \\
\cite{schweizer_shear_2008}     & PS      & 206~kDa    & 1.06      & 160\degree~\!\!C, 170\degree~\!\!C, 180\degree~\!\!C & 12.1 \\
\cite{bmk-10}             & UA-PE   & 400 CH$_2$ & 1         & 450~K        & 6    \\
\cite{sek-16}                   & UA-PE   & 700 CH$_2$ & 1         & 450~K        & 8.6  \\ \hline\hline
\end{tabular}
\caption{Normal stress ratio data sources and properties}
\label{table:N2N1}
\end{table}

\FloatBarrier

\bibliography{combined}
\end{document}